\pdfoutput=1
\documentclass[twocolumn,
               showpacs,
               preprintnumbers,
               prl,
               superscriptaddress,
               10pt,
               notitlepage,
               footinbib,
               aps]{revtex4-2}
               
\usepackage{graphicx,amssymb,amsmath,amsthm,amsfonts,epsfig,mathtools}

\usepackage[utf8]{inputenc}
\usepackage{graphicx}
\usepackage{float}
\usepackage{dcolumn}
\usepackage{bm}
\usepackage{color}
\usepackage{soul}
\usepackage[dvipsnames]{xcolor}
\usepackage{hyperref}
\hypersetup{colorlinks=true, citecolor=MidnightBlue,linkcolor=CornflowerBlue, urlcolor=CornflowerBlue, linktocpage=true}
\usepackage{xfrac}
\usepackage{siunitx}
\usepackage{empheq}
\usepackage[normalem]{ulem}

\newcommand{\beqn}{\begin{eqnarray}}
\newcommand{\eeqn}{\end{eqnarray}}
\newcommand{\beq}{\begin{equation}}
\newcommand{\eeq}{\end{equation}}
\newcommand{\citewithfootnote}[2]{%
  \begingroup
  \stepcounter{footnote}%
  \edef\currentfootnotekey{Note\thefootnote}%
  \footnotetext[\value{footnote}]{#2}%
  \expandafter\cite\expandafter{#1,\currentfootnotekey}%
  \endgroup
}

\def\mphi{m_{\phi}}

\def\rt{\tilde{\rho}}

\begin{document}

\title{Novel Observable Signals from First-Order Gravitational Phase Transitions}

\begin{abstract}
Spontaneous scalarization is a well-known phenomenon in scalar-tensor theories, featuring large deviations from general relativity. However, some of its signatures are hard to detect, because it is generally studied as a continuous phase transition. We show that scalarization occurs in a discontinuous (first-order) manner in many astrophysically relevant scenarios. This leads to loud signals that provide new tests of gravity. We demonstrate this in the case of gravitational wave breathing modes from a neutron star that scalarizes due to accretion, and discuss other possibilities.
\end{abstract}

\author{Murat \"Ozinan}
\email{mozinan21@ku.edu.tr}
\affiliation{Department of Physics, Ko\c{c} University, Rumelifeneri Yolu,\\
34450 Sar{\i}yer, \.{I}stanbul, T\"{u}rkiye}

\author{O\u{g}uzhan K. Yamak}
\email{oyamak21@ku.edu.tr}
\affiliation{Department of Physics, Ko\c{c} University, Rumelifeneri Yolu,\\
34450 Sar{\i}yer, \.{I}stanbul, T\"{u}rkiye}

\author{Fethi M. Ramazano\u{g}lu}
\email{framazanoglu@ku.edu.tr}
\affiliation{Department of Physics, Ko\c{c} University, Rumelifeneri Yolu,\\
34450 Sar{\i}yer, \.{I}stanbul, T\"{u}rkiye}

\date{\today}
\maketitle

\label{sec:introduction}

General relativity (GR) has successfully described gravitational phenomena over a wide range of length and energy scales ~\cite{Will:2014kxa,Barack:2018yly,GWTC-4-1, GWTC-4-2, GWTC-4-3}. Nevertheless, the strong-field and dynamical regime remains comparatively weakly tested, leaving open the possibility that additional gravitational degrees of freedom become active in the vicinity of compact objects. Scalar-tensor theories provide one of the simplest frameworks in which to investigate such deviations ~\cite{Fujii:2003pa}. In suitable regions of their parameter space, these theories can remain arbitrarily close to GR in weak gravitational fields, but predict nonperturbative departures from it under strong gravity~\cite{Doneva:2022ewd}.

The best-known example of this behavior is \textit{spontaneous scalarization}~\cite{Damour:1993hw,Damour:1992we,Damour:1996ke,Doneva:2022ewd}, or \textit{scalarization} in short. Above a critical compactness, the (approximately) vanishing scalar-field configuration develops a tachyonic instability, and the compact object acquires a macroscopic scalar cloud. The resulting scalarized configuration can differ significantly from its GR counterpart, despite the theory remaining nearly indistinguishable from GR in weak-field environments. This mechanism has motivated extensive studies of equilibrium and binary systems of neutron stars and black holes, stellar collapse, and their associated gravitational wave signatures ~\cite{Doneva:2022ewd, Sperhake:2017itk, Kuan:2022oxs, Unluturk:2025zie, Lam:2025ason, Kuan:2021,Silva:2018ssb, Herdeiro:2018, Li:2026}. 

Scalarization has traditionally been interpreted in analogy with spontaneous magnetization and described as a second-order phase transition~\cite{Damour:1996ke,Harada:1998ge}. In this picture, the scalarized branch emerges continuously from the GR branch in a stable manner, and the scalar field grows continuously from zero as the stellar mass crosses a critical value. Recent work has shown, however, that this continuous picture can be misleading. For example, the onset of scalarization in the original Damour and Esposito-Far\`ese (DEF) model of scalarization and its simple extensions feature a first-order phase transition in most parts of the parameter space~\cite{Unluturk:2025zie}: Arbitrarily weakly scalarized stars are unstable, the stable configurations being disconnected from the GR branch. Moreover, locally stable scalarized and unscalarized configurations may coexist at the same baryon mass (in the case of neutron stars), with one configuration representing the global energy minimum and the other a metastable local minimum. A sufficiently large perturbation can drive the star across the intervening energy barrier, producing a discontinuous and nonperturbative change in the scalar field and in the stellar structure. This realization has prompted investigations of first-order scalarization across a wide range of scalarization models~\cite{Unluturk:2025zie, Muniz:2025egq, Ozinan:2026bne, Huang:2025dgc, Guo:2025flg, Herdeiro:2026}. 

Our main aim here is demonstrating that the discontinuous nature of first-order scalarization considerably broadens the number of possible observation channels, and provides novel ways to test deviations from GR. In first-order scalarization, transitions between scalarized and unscalarized configurations are rapid, and the two structures are nonperturbatively different, unlike second-order scalarization. For example, in the case of neutron stars, such a transition can excite both scalar and fluid oscillations, and release part of the difference in binding energy between the two equilibrium configurations, leading to distinct observational signatures, whereas a continuous transition generally lacks such outcomes. This can be generalized to any dynamical study of scalarization in the literature in which a transition between scalarized and unscalarized configurations has been investigated; such studies have near-exclusively considered second-order (continuous) scalarization so far~\citewithfootnote{PJNee:2025, Julie:2023ncq, Barausse:2012da, Shibata:2013pra, Palenzuela:2013hsa, Dima:2020yac, Herdeiro:2020wei, Elley:2022ept, Pardoe:2026tcg, Capuano:2026, Elley:2022}{One important exception is \textcite{Kuan:2022oxs}, which considered a discontinuous transition at the maximum mass limit of stellar solutions, a distinct phenomenon from the later discover first-order scalarization picture we study here.}.

There has been one obstacle to the above scenario: first-order scalarization typically occurs at lower, $\lesssim 1M_\odot$, stellar masses, if one uses the original, and arguably the simplest, scalarization model of DEF~\cite{Unluturk:2025zie,Ozinan:2026bne}. This does not change the phenomenology, but lowers the observational prospects in our universe, where neutron stars are typically more massive~\cite{Ozel:2016oaf}.

We show that low stellar masses is not an impediment to observe first-order scalarization, but rather a restriction arising from assuming a particular form of scalar-tensor theory coupling. Namely, the order of scalarization is governed by the quartic (fourth-order) coupling between the scalar field and matter, which is directly related to the quadratic (second-order) coupling constant $\beta$ in the DEF model. This relationship is not due to any specific physical reasoning to the best of our knowledge, but has been followed for simplicity of investigation in most existing work. When the whole available range for the quartic coupling is considered (without resorting to fine tuning), the onset of first-order scalarization occurs at much broader stellar masses, including the most astrophysically relevant ones. This was first hinted in the context of equilibrium neutron stars~\cite{Ozinan:2026bne}, and we will show how it leads to first-order scalarization becoming manifestly relevant for compact object astronomy in dynamical systems.

We mentioned that considering first-order scalarization rather than second-order brings new phenomenology to many existing studies of scalarization, which form a vast literature~\cite{Doneva:2022ewd}. Our initial study will focus on one particularly interesting case: gravitational wave breathing modes arising from spherically symmetric oscillations of a neutron star. Starting from an equilibrium configuration, we simulate a model accretion that drives a neutron star across the energy barrier, and follow its nonlinear migration to a strongly scalarized branch via first-order scalarization. We compute the accompanying changes in the stellar structure, the scalar radiation emitted during the transition, and the subsequent propagation of the massive scalar signal to astrophysical distances, leading to a gravitational breathing mode. 

Our scalar field is massive, which features additional interesting observations due to dispersion~\cite{Sperhake:2017itk}. A short burst emitted at the source is stretched during propagation into a long-lived inverse-chirp signal, with the high-frequency components arriving first at far-away observers~\cite{Sperhake:2017itk,Kuan:2022oxs}. We intend this to be an initial attempt at demonstrating the many observational possibilities offered by first-order gravitational transitions, which is to be followed by more involved studies in binary systems.

\noindent {\bf \em   Spontaneous scalarization:}
\label{sec:theory}
We briefly summarize neutron star scalarization in the original DEF model~\cite{Damour:1993hw} for a general audience. The essential points here can be extended to generic scalar-tensor theories~\cite{Doneva:2017bvd,Silva:2017uqg,Andreou:2019ikc,Doneva:2022ewd}. We use geometrized units $G=c=\hbar =1$ unless otherwise stated.

Consider the scalar-tensor theory action
\begin{align}
S =
&\frac{1}{16\pi }
\int d^4x\,\sqrt{-g}\,
\left[
R
-2g^{\mu\nu}\nabla_\mu\varphi\nabla_\nu\varphi
-4V(\varphi)
\right]
\nonumber \\
&+S_{\rm m}
\left[
\Psi_{\rm m};
\tilde{g}_{\mu\nu}
\right].
\label{eq:action}
\end{align}
Here, $g_{\mu\nu}$ is the so-called Einstein-frame metric, $R$ is its Ricci scalar, and $\varphi$ is the scalar field. The matter fields, collectively denoted by $\Psi_{\rm m}$, couple minimally to $\tilde{g}_{\mu\nu}=A^2(\varphi)g_{\mu\nu}$, which is called the Jordan-frame metric. We use $V(\varphi)=\sfrac{1}{2}\ m_\varphi^2\varphi^2$ corresponding to a massive scalar.

Variation of Eq.~\eqref{eq:action} gives the scalar field equation
\begin{align}
\Box_g\varphi
={}&
-4\pi\alpha(\varphi)T
+
m_\varphi^2\varphi,
\label{eq:scalar_equation}
\end{align}
where $\alpha(\varphi) \equiv \tfrac{d\ln A(\varphi)}{d\varphi},$ $T_{\mu\nu}$ is the Einstein-frame matter stress-energy tensor, and $T=g^{\mu\nu}T_{\mu\nu}$ is its trace. For the stellar matter, we assume a perfect fluid.

We adopt the conformal coupling
\begin{equation}
A(\varphi)
=
\exp\left(
\alpha_0\varphi
+
\frac{1}{2}\beta\varphi^2
-
\frac{1}{24}\gamma\varphi^4
\right).
\label{eq:conformal_factor}
\end{equation}
The linear coefficient $\alpha_0$ is not essential for the existence of scalarization, but it has an important role in observational relevance we will elaborate on later. The quadratic coefficient $\beta$ is the one mainly responsible for the usual tachyonic scalarization mechanism, while the quartic term $\gamma$ has a dominant effect on the order of the phase transition~\citewithfootnote{Unluturk:2025zie,Ozinan:2026bne}{Similar functions have been used in the context of nonlinear scalarization as well~\cite{Doneva:2021tvn}.}.

\noindent {\bf \em The phase transition structure:}
The distinction between continuous and discontinuous scalarization can be expressed through a Landau-type energy expansion as detailed in \textcite{Unluturk:2025zie}. In the symmetric $\alpha_0=0$ limit, the ADM mass can be expressed as
\begin{equation}
M_{\rm ADM}
=
M_0
+
a\Phi^2
+
\frac{1}{2}b\Phi^4
+
\frac{1}{3}c\Phi^6
+\cdots ,
\label{eq:landau}
\end{equation}
where $\Phi$ is some measure of the scalarization strength, each coefficient depends on the baryon mass $M_{\rm b}$, and $c>0$ ensures that the energy is bounded from below. For $b > 0$, scalarized minima emerge continuously when $a$ changes sign, corresponding to second-order scalarization. By contrast, $b<0$ permits locally stable scalarized and unscalarized minima to coexist while $a$ is still positive. The system then contains an energy barrier separating the two phases. The globally preferred configuration changes discontinuously at the point where the two minima have equal energy, producing a first-order transition. Since the coefficients in Eq.~\eqref{eq:landau} are baryon mass-dependent, the transition can be triggered by changes in the matter content of the star.

The sign of $b$ that controls the order of scalarization, first or second, arises from the form of $A(\varphi)$ as well as the response of the stellar matter. The exact form of $A(\varphi)$ is not essential; the $\alpha_0, \beta, \gamma$ parameters we employ are only relevant in terms of their contributions to the Landau energy~\eqref{eq:landau}. In a sense, the exponential form in Eq.~\eqref{eq:conformal_factor} obscures the actual physical coupling between matter and the scalar field, which is more apparent in the Taylor expansion of $A^4(\varphi)$~\cite{Ozinan:2026bne}. With this physical coupling in mind, positive $\gamma$ values are known to facilitate first-order scalarization by decreasing $b$, and we will study $\gamma = 12\beta^2$ unless stated otherwise. This is because the aforementioned Taylor expansion has contributions from both $\beta$ and $\gamma$ at the quartic ($\varphi^4$) order, and $\gamma = 12\beta^2$ exactly ``neutralizes'' this contribution, representing a natural case to investigate: $A^4(\varphi) = \dots + \left( 12\beta^2-\gamma\right) \varphi^4/6+\dots$ ($\alpha_0=0$). We study other values as well to demonstrate that our results do not rely on fine tuning $A(\varphi)$~\footnote{We could easily construct a single parameter family of functions that has a vanishing quadratic term in the Taylor expansion for $A^4$, such as $A_\rho^4(\varphi) = 1 + 1/2 \tanh {(4\rho\varphi^2)}$, which would prefer first-order scalarization similarly to Eq.~\eqref{eq:conformal_factor} with $\gamma = 12\beta^2$. This would explicitly demonstrate that the number of parameters in $A(\varphi)$ is not a direct measure of how complicated or fine tuned the theory is. However we chose to keep the current form in Eq.~\eqref{eq:conformal_factor} that has the additional $\gamma$ parameter for ease of comparison to the existing literature.}.

A finite $\alpha_0$ breaks the exact $\varphi\rightarrow-\varphi$ symmetry and weakly tilts the energy landscape. Consequently, the unscalarized solutions are now weakly scalarized as opposed to strongly scalarized solutions that are deformed versions of the original scalarized solutions for $\alpha_0 = 0$. Nevertheless, for sufficiently small $\alpha_0$, the characteristics of first-order scalarization persist, and we still refer to the transition between the weakly and strongly scalarized configurations as scalarization~\cite{Staykov:2026}. $\alpha_0$ contributions at the quartic order we discussed are negligible for commonly considered values, so the effect of $\gamma$ is essentially the same.

\begin{figure}
    \centering
    \includegraphics[width=\columnwidth]{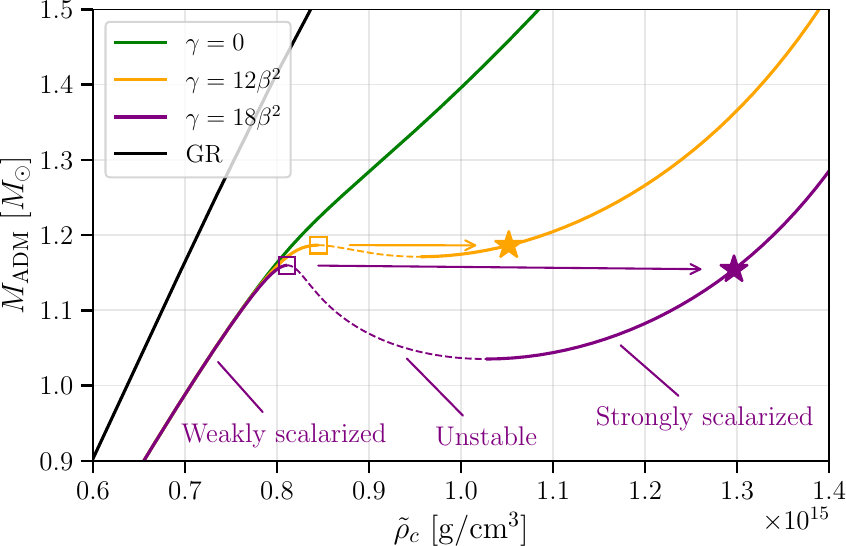}
    \caption{Baryon mass $M_b$ as a function of the central energy density $\rt_c$ for equilibrium neutron-star configurations. The GR sequence is compared with scalar-tensor models with $\beta=-5, \, \mphi = 10^{-14} \text{ eV}$ and $\gamma = 0, \, 12\beta^2, \, 18\beta^2 $. Increasing $\gamma$ facilitates first-order scalarization as indicated by the presence of two distinct locally stable solutions at certain stellar masses. Stellar accretion triggers the first-order phase transitions shown with the arrows.}
   
    \label{fig: branch}
\end{figure}
%
\noindent {\bf \em   Results:}
\label{sec:results_discussion}
Aside from the different values for $\gamma$, we use theory parameters commonly considered in the literature unless stated otherwise: $\alpha = 10^{-2}, \, \beta = -5, \, \mphi = 10^{-14} \text{ eV}$ \cite{Sperhake:2017itk, Rosca-Mead:2020bzt, Kuan:2022oxs}. We will comment on the effect of different choices after presenting our results. As the phase transition structure is not EOS sensitive \cite{Ozinan:2026bne}, we only consider HB EOS~\cite{Read:2009yp} as a representative case.

Equilibrium solutions for first- and second-order scalarized neutron stars for these parameters can be seen in Figure~\ref{fig: branch}. Particularly note that as we follow stars with increasing masses, the stable solutions, discontinuously jump once we go above a local maximum, marked by squares. This is the scenario we will be examining, which naturally occurs for an accreting star. The stable solutions to the left of this point are weakly scalarized (they are not GR configurations due to $\alpha_0 \neq 0$), and the ones farther to the right are strongly scalarized. In between, we have unstable solutions connecting the two~\cite{Unluturk:2025zie,Staykov:2026}.

In our radially symmetric numerical computation, we select the last equilibrium star on the weakly scalarized branch (the {square} in Figure~\ref{fig: branch}) as the initial configuration. We then approximate the final stage of accretion by introducing a spherically symmetric Gaussian shell of matter around the neutron star, motivated by the accretion prescription of \textcite{Kuan:2022oxs}. The absorption of this blob pushes the star above the allowed maximum stellar mass (locally), triggering the transition we mentioned, which is shown with arrows in Figure~\ref{fig: branch}. This ``toy model'' of accretion is sufficient to observe the transition as has been previously demonstrated~\cite{Kuan:2022oxs}, and the details of our numerical computations are in the ``End Matter.'' 

\begin{figure}
\centering
\includegraphics[width=\columnwidth]{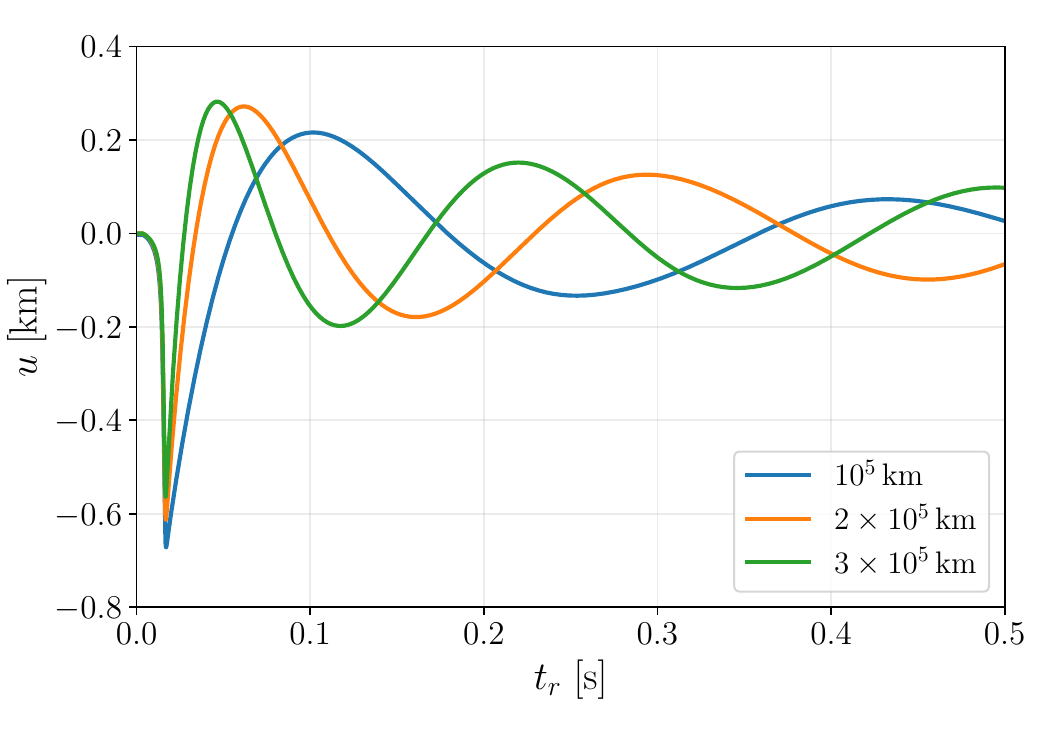}
\caption{Scalar radiation for the $\gamma = 12 \beta^2$ transition extracted at different radii. This shows the dispersion and the resulting earlier arrival of higher frequency modes, leading to an inverse chirp for far-away observers in Figure~\ref{fig:sensitivity}.}
\label{fig:inverse_chirp}
\end{figure}

The transition process takes place in the dynamical time frame of the stellar oscillations. It involves the abrupt growth of the scalar field, the contraction of the star and the increase in the density at its center. The remnant subsequently undergoes coupled hydrodynamic and scalar oscillations about the new equilibrium configuration. This generates a strong monopolar burst of scalar and gravitational radiation, reflecting the energy difference between the ADM masses of the weakly and strongly scalarized phases, i.e., the latent heat associated with the first-order phase transition. The scalar field radiation is presented in Figure~\ref{fig:inverse_chirp}.

We should note that we are primarily interested in the gravitational wave detections from such an event, not the scalar waves. No gravitational waves are produced under spherical symmetry in GR, but there are breathing and longitudinal modes in scalar-tensor theories. The former mode has the strain
\begin{equation}\label{eq:breathing}
    h_{\rm B} = A^2(\varphi)-1 \approx 2\alpha_0 \varphi ,    
\end{equation}
and the latter signal is smaller by a factor of $(f_*/f)^2$, $f$ being the wave frequency and $f_*=\mphi/2\pi=2.42\,\mathrm{Hz}$ the frequency corresponding to the scalar mass~\cite{Will:2014kxa,Sperhake:2017itk,Kuan:2021}. Also see the ``End Matter.'' Overall, the scalar waves directly govern our gravitational wave signals, hence our emphasis on their extraction.

Because the scalar field is massive, the emitted frequency components propagate with different group velocities $v_g = \sqrt{1-(f_*/f)^2}$, with frequencies below $f_*$ decaying exponentially. High-frequency modes travel close to the speed of light and arrive first, whereas lower-frequency modes experience progressively larger delays~\cite{Sperhake:2017itk}. As shown in Figure~\ref{fig:inverse_chirp}, this dispersion converts the short source burst into a long-lived inverse chirp whose instantaneous frequency decreases with time.

\begin{figure}
\centering
\includegraphics[width=\columnwidth]{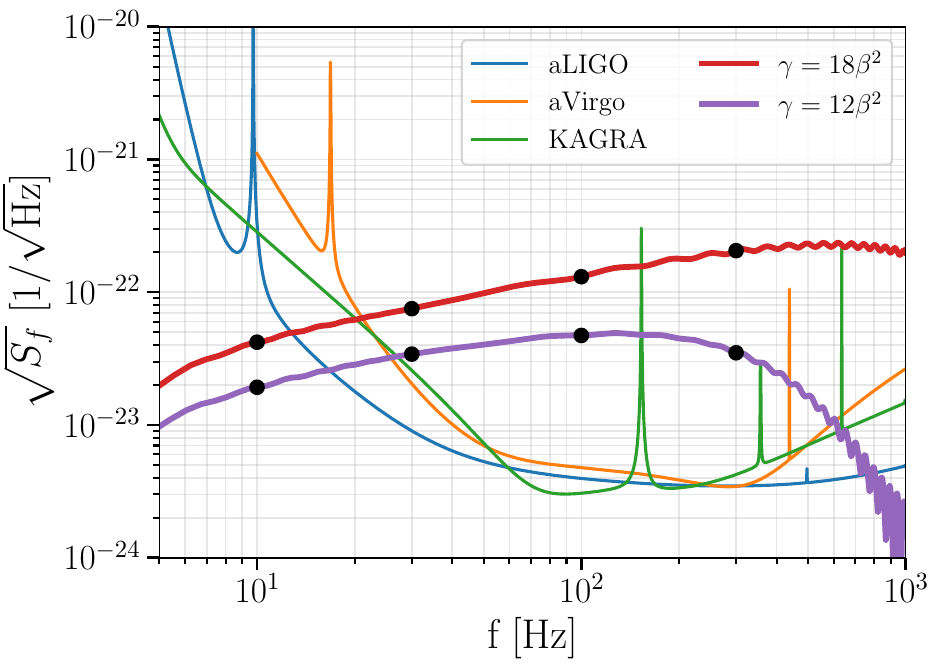}
\caption{Effective amplitude spectral density, $\sqrt{S_f}$, of the breathing mode for the two coupling choices $\gamma = 12\beta^2$ and $\gamma = 18\beta^2$ at a source distance of $D=10\,\mathrm{kpc}$, assuming a coherent observation time of $T=60$ days, compared with representative detector noise amplitude spectral densities~\cite{LIGO-T1500293}. $\gamma=0$, leading to second-order scalarization, does not produce a signal distinguishable from numerical error. The black markers indicate the quasimonochromatic signal frequency at selected retarded times; from right to left, they correspond to $t=1$, $10$, $100$, and $1000$ years after the transition.}
\label{fig:sensitivity}
\end{figure}
The dispersion is not a mild one for typical parameter values such as the ones we use. The signal is so stretched that far away observers detect an almost-constant frequency at any moment, which slowly shifts on timescales of years, for example, for a source at a distance $D=10$kpc. More precisely, the observed frequency is~\cite{Sperhake:2017itk}
\begin{equation}\label{eq:frequency}
    f(t) \approx f_*/\sqrt{1-(D/t)^2} 
    \approx f_*/\sqrt{2(t-D)/D} ,    
\end{equation}
$t>D$ being the time since the source event, and the last approximation being valid in the regime $(t-D) \ll D$. We depict this in Figure~\ref{fig:sensitivity}, where the frequency dependence of the strain curve corresponds to signals extracted at a relatively close point to the source. The amplitude is scaled for the observation distance and a coherent observation time. As explained, the far away observer measures a single point corresponding to a single frequency on the strain curve at any given time. Thus, specialized narrow-band searches would be especially fruitful to capture this signal~\cite{Kuan:2022oxs}.

Most crucially for our purposes, Figure~\ref{fig:sensitivity} also shows what happens as the value of $\gamma$ changes. Increasing this parameter typically increases the structural differences between the endpoints of the scalarization transition, hence enhances the signal. However, the strain is high in all cases where there is a first-order transition. Scalarization is second-order (continuous) when $\gamma=0$ for our parameters, which leads to no abrupt configuration change of the star, hence no significant signal. This contrast between the first- and second-order scalarization is the main point of this example. We reiterate that $\gamma=12\beta^2$, rather than $\gamma=0$, is the ``natural'' value of this parameter, and any higher value makes our findings stronger.

As for the dependence of the signal on the other parameters, $\alpha_0$ is not essential for the occurrence of a discontinuous transition or the main features of the stellar oscillations. However, recall that the gravitational wave strain in our spherically symmetric system is proportional to $\alpha_0$ in Eq.~\eqref{eq:breathing}, hence its existence and value are crucial for observability. 

Making $\beta$ more negative while keeping all other parameters constant enhances first-order scalarization, but doing so excessively also lowers the stellar masses where the transition occurs, hence is undesirable. With all else kept the same, the stellar mass at the transition becomes $1M_\odot$ at $\beta \approx -5.8$ for our EOS. Finally, increasing $\mphi$ suppresses scalarization and moves us towards second-order scalarization, but also increases the transition stellar mass, so its effects on the transition we studied are not straightforward~\cite{Ramazanoglu:2016kul,Tuna:2022qqr,Ozinan:2026bne}. 

Perhaps the most important consequence of $\mphi$ is the linear scaling of the observed frequency in Eq.~\eqref{eq:frequency}, so that the signal could completely leave the frequency range of the current detectors when $\mphi \gtrsim 10^{-11}\, \mathrm{eV}$. Some observational considerations suggest the constraint $m_\varphi\gtrsim10^{-11}\,{\rm eV}$ for $\alpha_0=0=\gamma$~\cite{Kuan:2023hrh}, but these limits are qualitative and do not yet consider the first-order scalarization that we study. 

To summarize, the observability of our computed signal can change significantly with the theory parameters. It is sometimes even more enhanced and sometimes diminished, but a large part of the parameter space leads to the loud signals we are interested in.

\noindent {\bf \em   Discussion:}
\label{sec:results_discussion}
We will not delve into the further details of the specific scalarization transition we studied, since that is not the main aim of this manuscript, and they have been worked out in related studies we mentioned~\cite{Gerosa:2016fri, Sperhake:2017itk, Rosca-Mead:2020bzt,Kuan:2022oxs}. Rather, our main result is the general fact that when second-order scalarization is replaced by its first-order version, astrophysical phenomena and the related observational signals are radically altered due to the discontinuous nature of the latter, leading to novel signatures for deviations from GR. We used a transition from a weakly scalarized neutron star configuration to a strongly scalarized one as a representative example, partially due to the relative tractability of the numerics under spherical symmetry. However, this is just a specific example, and the main take-away can easily be adapted to any scenario of scalarization or descalarization.

One direction for generalization is the scalarization of black holes~\cite{Doneva:2017bvd,Silva:2017uqg,Andreou:2019ikc,Doneva:2022ewd}, where discontinuous transitions have only been investigated in limited models and static settings~\cite{Herdeiro:2026}. Black hole systems also have the potential for further observables thanks to the fact that (de)scalarization can be strongly triggered by changing their spin as well as their mass~\cite{Dima:2020yac,Herdeiro:2020wei,Berti:2020kgk}. A spin effect can be a new channel for neutron stars as well, but it is relatively weaker~\cite{Doneva:2013qva}. Black holes also provide an independent check on any possible signal since they present a lower risk of a first-order \emph{nuclear phase transition} being misinterpreted as scalarization~\cite{Sagert:2009, Shao:2019gjj, Most:2019}, though this may not be a major issue due to different stellar mass scales involved~\cite{Cao:2026dwk}.

Binaries promise some of the most striking observables from (de)scalarization. For example, an external scalar field can facilitate scalarization in general. Hence, an object in a binary, one which would not scalarize otherwise, can feature \emph{induced scalarization} due to the scalar field of its already scalarized companion, when the orbital separation becomes small enough~\cite{Salgado:1998sg,Barausse:2012da}. In \emph{dynamical scalarization,} the phase transition is triggered by an effective total compactness when the separation becomes small in a binary, even though the objects do not scalarize in isolation~\cite{Palenzuela:2013hsa,Shibata:2013pra,Taniguchi:2014fqa,Khalil:2019wyy}. For these two phenomena, scalarization is the natural transition in quasi-circular binaries, but descalarization is also possible on eccentric orbits. The part crucial for our discussion is that these processes occur gradually in second-order scalarization, but they would lead to abrupt changes and distinct observables when a first-order transition occurs. 

We discussed a limited number of the dynamical phenomena in which we expect first-order scalarization to have a drastic effect. However, the vast literature of scalarization, now in its fourth decade, will likely provide many other examples. As the order of scalarization is directly related to the quartic coupling of the scalar field to gravity and matter, these observations in turn can provide a means to measure or constrain it.

\acknowledgments
We thank K{\i}van\c{c} \.I. \"Unl\"ut\"urk, Takami Kuroda and Masaru Shibata for their comments. F.M.R was supported by the Scientific and Technological Research Council of Turkey (T\"UB\.ITAK) Grant Number 122F097.

\bibliography{references_all}

\begin{thebibliography}{64}%
\makeatletter
\providecommand \@ifxundefined [1]{%
 \@ifx{#1\undefined}
}%
\providecommand \@ifnum [1]{%
 \ifnum #1\expandafter \@firstoftwo
 \else \expandafter \@secondoftwo
 \fi
}%
\providecommand \@ifx [1]{%
 \ifx #1\expandafter \@firstoftwo
 \else \expandafter \@secondoftwo
 \fi
}%
\providecommand \natexlab [1]{#1}%
\providecommand \enquote  [1]{``#1''}%
\providecommand \bibnamefont  [1]{#1}%
\providecommand \bibfnamefont [1]{#1}%
\providecommand \citenamefont [1]{#1}%
\providecommand \href@noop [0]{\@secondoftwo}%
\providecommand \href [0]{\begingroup \@sanitize@url \@href}%
\providecommand \@href[1]{\@@startlink{#1}\@@href}%
\providecommand \@@href[1]{\endgroup#1\@@endlink}%
\providecommand \@sanitize@url [0]{\catcode `\\12\catcode `\$12\catcode `\&12\catcode `\#12\catcode `\^12\catcode `\_12\catcode `\%12\relax}%
\providecommand \@@startlink[1]{}%
\providecommand \@@endlink[0]{}%
\providecommand \url  [0]{\begingroup\@sanitize@url \@url }%
\providecommand \@url [1]{\endgroup\@href {#1}{\urlprefix }}%
\providecommand \urlprefix  [0]{URL }%
\providecommand \Eprint [0]{\href }%
\providecommand \doibase [0]{https://doi.org/}%
\providecommand \selectlanguage [0]{\@gobble}%
\providecommand \bibinfo  [0]{\@secondoftwo}%
\providecommand \bibfield  [0]{\@secondoftwo}%
\providecommand \translation [1]{[#1]}%
\providecommand \BibitemOpen [0]{}%
\providecommand \bibitemStop [0]{}%
\providecommand \bibitemNoStop [0]{.\EOS\space}%
\providecommand \EOS [0]{\spacefactor3000\relax}%
\providecommand \BibitemShut  [1]{\csname bibitem#1\endcsname}%
\let\auto@bib@innerbib\@empty
\bibitem [{\citenamefont {Will}(2014)}]{Will:2014kxa}%
  \BibitemOpen
  \bibfield  {author} {\bibinfo {author} {\bibfnamefont {C.~M.}\ \bibnamefont {Will}},\ }\bibfield  {title} {\bibinfo {title} {{The Confrontation between General Relativity and Experiment}},\ }\href {https://doi.org/10.12942/lrr-2014-4} {\bibfield  {journal} {\bibinfo  {journal} {Living Rev. Rel.}\ }\textbf {\bibinfo {volume} {17}},\ \bibinfo {pages} {4} (\bibinfo {year} {2014})},\ \Eprint {https://arxiv.org/abs/1403.7377} {arXiv:1403.7377 [gr-qc]} \BibitemShut {NoStop}%
\bibitem [{\citenamefont {Barack}\ \emph {et~al.}(2019)\citenamefont {Barack} \emph {et~al.}}]{Barack:2018yly}%
  \BibitemOpen
  \bibfield  {author} {\bibinfo {author} {\bibfnamefont {L.}~\bibnamefont {Barack}} \emph {et~al.},\ }\bibfield  {title} {\bibinfo {title} {{Black holes, gravitational waves and fundamental physics: a roadmap}},\ }\href {https://doi.org/10.1088/1361-6382/ab0587} {\bibfield  {journal} {\bibinfo  {journal} {Class. Quant. Grav.}\ }\textbf {\bibinfo {volume} {36}},\ \bibinfo {pages} {143001} (\bibinfo {year} {2019})},\ \Eprint {https://arxiv.org/abs/1806.05195} {arXiv:1806.05195 [gr-qc]} \BibitemShut {NoStop}%
\bibitem [{\citenamefont {Collaboration}\ \emph {et~al.}(2026{\natexlab{a}})\citenamefont {Collaboration}, \citenamefont {the Virgo~Collaboration},\ and\ \citenamefont {the KAGRA~Collaboration}}]{GWTC-4-1}%
  \BibitemOpen
  \bibfield  {author} {\bibinfo {author} {\bibfnamefont {T.~L.~S.}\ \bibnamefont {Collaboration}}, \bibinfo {author} {\bibnamefont {the Virgo~Collaboration}},\ and\ \bibinfo {author} {\bibnamefont {the KAGRA~Collaboration}},\ }\href {https://arxiv.org/abs/2603.19019} {\bibinfo {title} {Gwtc-4.0: Tests of general relativity. i. overview and general tests}} (\bibinfo {year} {2026}{\natexlab{a}}),\ \Eprint {https://arxiv.org/abs/2603.19019} {arXiv:2603.19019 [gr-qc]} \BibitemShut {NoStop}%
\bibitem [{\citenamefont {Collaboration}\ \emph {et~al.}(2026{\natexlab{b}})\citenamefont {Collaboration}, \citenamefont {the Virgo~Collaboration},\ and\ \citenamefont {the KAGRA~Collaboration}}]{GWTC-4-2}%
  \BibitemOpen
  \bibfield  {author} {\bibinfo {author} {\bibfnamefont {T.~L.~S.}\ \bibnamefont {Collaboration}}, \bibinfo {author} {\bibnamefont {the Virgo~Collaboration}},\ and\ \bibinfo {author} {\bibnamefont {the KAGRA~Collaboration}},\ }\href {https://arxiv.org/abs/2603.19020} {\bibinfo {title} {Gwtc-4.0: Tests of general relativity. ii. parameterized tests}} (\bibinfo {year} {2026}{\natexlab{b}}),\ \Eprint {https://arxiv.org/abs/2603.19020} {arXiv:2603.19020 [gr-qc]} \BibitemShut {NoStop}%
\bibitem [{\citenamefont {Collaboration}\ \emph {et~al.}(2026{\natexlab{c}})\citenamefont {Collaboration}, \citenamefont {the Virgo~Collaboration},\ and\ \citenamefont {the KAGRA~Collaboration}}]{GWTC-4-3}%
  \BibitemOpen
  \bibfield  {author} {\bibinfo {author} {\bibfnamefont {T.~L.~S.}\ \bibnamefont {Collaboration}}, \bibinfo {author} {\bibnamefont {the Virgo~Collaboration}},\ and\ \bibinfo {author} {\bibnamefont {the KAGRA~Collaboration}},\ }\href {https://arxiv.org/abs/2603.19021} {\bibinfo {title} {Gwtc-4.0: Tests of general relativity. iii. tests of the remnants}} (\bibinfo {year} {2026}{\natexlab{c}}),\ \Eprint {https://arxiv.org/abs/2603.19021} {arXiv:2603.19021 [gr-qc]} \BibitemShut {NoStop}%
\bibitem [{\citenamefont {Fujii}\ and\ \citenamefont {Maeda}(2003)}]{Fujii:2003pa}%
  \BibitemOpen
  \bibfield  {author} {\bibinfo {author} {\bibfnamefont {Y.}~\bibnamefont {Fujii}}\ and\ \bibinfo {author} {\bibfnamefont {K.}~\bibnamefont {Maeda}},\ }\href@noop {} {\emph {\bibinfo {title} {{The Scalar-Tensor Theory of Gravitation}}}}\ (\bibinfo  {publisher} {Cambridge University Press},\ \bibinfo {address} {Cambridge},\ \bibinfo {year} {2003})\BibitemShut {NoStop}%
\bibitem [{\citenamefont {Doneva}\ \emph {et~al.}(2024)\citenamefont {Doneva}, \citenamefont {Ramazano\u{g}lu}, \citenamefont {Silva}, \citenamefont {Sotiriou},\ and\ \citenamefont {Yazadjiev}}]{Doneva:2022ewd}%
  \BibitemOpen
  \bibfield  {author} {\bibinfo {author} {\bibfnamefont {D.~D.}\ \bibnamefont {Doneva}}, \bibinfo {author} {\bibfnamefont {F.~M.}\ \bibnamefont {Ramazano\u{g}lu}}, \bibinfo {author} {\bibfnamefont {H.~O.}\ \bibnamefont {Silva}}, \bibinfo {author} {\bibfnamefont {T.~P.}\ \bibnamefont {Sotiriou}},\ and\ \bibinfo {author} {\bibfnamefont {S.~S.}\ \bibnamefont {Yazadjiev}},\ }\bibfield  {title} {\bibinfo {title} {{Spontaneous scalarization}},\ }\href {https://doi.org/10.1103/RevModPhys.96.015004} {\bibfield  {journal} {\bibinfo  {journal} {Rev. Mod. Phys.}\ }\textbf {\bibinfo {volume} {96}},\ \bibinfo {pages} {015004} (\bibinfo {year} {2024})},\ \Eprint {https://arxiv.org/abs/2211.01766} {arXiv:2211.01766 [gr-qc]} \BibitemShut {NoStop}%
\bibitem [{\citenamefont {Damour}\ and\ \citenamefont {Esposito-Far\`ese}(1993)}]{Damour:1993hw}%
  \BibitemOpen
  \bibfield  {author} {\bibinfo {author} {\bibfnamefont {T.}~\bibnamefont {Damour}}\ and\ \bibinfo {author} {\bibfnamefont {G.}~\bibnamefont {Esposito-Far\`ese}},\ }\bibfield  {title} {\bibinfo {title} {{Nonperturbative strong-field effects in tensor-scalar theories of gravitation}},\ }\href {https://doi.org/10.1103/PhysRevLett.70.2220} {\bibfield  {journal} {\bibinfo  {journal} {Phys. Rev. Lett.}\ }\textbf {\bibinfo {volume} {70}},\ \bibinfo {pages} {2220} (\bibinfo {year} {1993})}\BibitemShut {NoStop}%
\bibitem [{\citenamefont {Damour}\ and\ \citenamefont {Esposito-Far{\`e}se}(1992)}]{Damour:1992we}%
  \BibitemOpen
  \bibfield  {author} {\bibinfo {author} {\bibfnamefont {T.}~\bibnamefont {Damour}}\ and\ \bibinfo {author} {\bibfnamefont {G.}~\bibnamefont {Esposito-Far{\`e}se}},\ }\bibfield  {title} {\bibinfo {title} {{Tensor multiscalar theories of gravitation}},\ }\href {https://doi.org/10.1088/0264-9381/9/9/015} {\bibfield  {journal} {\bibinfo  {journal} {Class.Quant.Grav.}\ }\textbf {\bibinfo {volume} {9}},\ \bibinfo {pages} {2093} (\bibinfo {year} {1992})}\BibitemShut {NoStop}%
\bibitem [{\citenamefont {Damour}\ and\ \citenamefont {Esposito-Far{\`e}se}(1996)}]{Damour:1996ke}%
  \BibitemOpen
  \bibfield  {author} {\bibinfo {author} {\bibfnamefont {T.}~\bibnamefont {Damour}}\ and\ \bibinfo {author} {\bibfnamefont {G.}~\bibnamefont {Esposito-Far{\`e}se}},\ }\bibfield  {title} {\bibinfo {title} {{Tensor-scalar gravity and binary-pulsar experiments}},\ }\href {https://doi.org/10.1103/PhysRevD.54.1474} {\bibfield  {journal} {\bibinfo  {journal} {Phys. Rev. D}\ }\textbf {\bibinfo {volume} {54}},\ \bibinfo {pages} {1474} (\bibinfo {year} {1996})},\ \Eprint {https://arxiv.org/abs/gr-qc/9602056} {arXiv:gr-qc/9602056} \BibitemShut {NoStop}%
\bibitem [{\citenamefont {Sperhake}\ \emph {et~al.}(2017)\citenamefont {Sperhake}, \citenamefont {Moore}, \citenamefont {Rosca}, \citenamefont {Agathos}, \citenamefont {Gerosa},\ and\ \citenamefont {Ott}}]{Sperhake:2017itk}%
  \BibitemOpen
  \bibfield  {author} {\bibinfo {author} {\bibfnamefont {U.}~\bibnamefont {Sperhake}}, \bibinfo {author} {\bibfnamefont {C.~J.}\ \bibnamefont {Moore}}, \bibinfo {author} {\bibfnamefont {R.}~\bibnamefont {Rosca}}, \bibinfo {author} {\bibfnamefont {M.}~\bibnamefont {Agathos}}, \bibinfo {author} {\bibfnamefont {D.}~\bibnamefont {Gerosa}},\ and\ \bibinfo {author} {\bibfnamefont {C.~D.}\ \bibnamefont {Ott}},\ }\bibfield  {title} {\bibinfo {title} {{Long-lived inverse chirp signals from core collapse in massive scalar-tensor gravity}},\ }\href {https://doi.org/10.1103/PhysRevLett.119.201103} {\bibfield  {journal} {\bibinfo  {journal} {Phys. Rev. Lett.}\ }\textbf {\bibinfo {volume} {119}},\ \bibinfo {pages} {201103} (\bibinfo {year} {2017})},\ \Eprint {https://arxiv.org/abs/1708.03651} {arXiv:1708.03651 [gr-qc]} \BibitemShut {NoStop}%
\bibitem [{\citenamefont {Kuan}\ \emph {et~al.}(2022)\citenamefont {Kuan}, \citenamefont {Suvorov}, \citenamefont {Doneva},\ and\ \citenamefont {Yazadjiev}}]{Kuan:2022oxs}%
  \BibitemOpen
  \bibfield  {author} {\bibinfo {author} {\bibfnamefont {H.-J.}\ \bibnamefont {Kuan}}, \bibinfo {author} {\bibfnamefont {A.~G.}\ \bibnamefont {Suvorov}}, \bibinfo {author} {\bibfnamefont {D.~D.}\ \bibnamefont {Doneva}},\ and\ \bibinfo {author} {\bibfnamefont {S.~S.}\ \bibnamefont {Yazadjiev}},\ }\bibfield  {title} {\bibinfo {title} {{Gravitational Waves from Accretion-Induced Descalarization in Massive Scalar-Tensor Theory}},\ }\href {https://doi.org/10.1103/PhysRevLett.129.121104} {\bibfield  {journal} {\bibinfo  {journal} {Phys. Rev. Lett.}\ }\textbf {\bibinfo {volume} {129}},\ \bibinfo {pages} {121104} (\bibinfo {year} {2022})},\ \Eprint {https://arxiv.org/abs/2203.03672} {arXiv:2203.03672 [gr-qc]} \BibitemShut {NoStop}%
\bibitem [{\citenamefont {{\"U}nl{\"u}t{\"u}rk}\ \emph {et~al.}(2025)\citenamefont {{\"U}nl{\"u}t{\"u}rk}, \citenamefont {Tuna}, \citenamefont {Yamak},\ and\ \citenamefont {Ramazano{\u{g}}lu}}]{Unluturk:2025zie}%
  \BibitemOpen
  \bibfield  {author} {\bibinfo {author} {\bibfnamefont {K.~{\.I}.}\ \bibnamefont {{\"U}nl{\"u}t{\"u}rk}}, \bibinfo {author} {\bibfnamefont {S.}~\bibnamefont {Tuna}}, \bibinfo {author} {\bibfnamefont {O.~K.}\ \bibnamefont {Yamak}},\ and\ \bibinfo {author} {\bibfnamefont {F.~M.}\ \bibnamefont {Ramazano{\u{g}}lu}},\ }\bibfield  {title} {\bibinfo {title} {Nature of phase transitions and metastability in scalar-tensor theories},\ }\href {https://doi.org/10.1103/14xz-d7xf} {\bibfield  {journal} {\bibinfo  {journal} {Phys. Rev. Lett.}\ }\textbf {\bibinfo {volume} {135}},\ \bibinfo {pages} {061401} (\bibinfo {year} {2025})},\ \Eprint {https://arxiv.org/abs/2502.01781} {arXiv:2502.01781 [gr-qc]} \BibitemShut {NoStop}%
\bibitem [{\citenamefont {Lam}\ \emph {et~al.}(2025)\citenamefont {Lam}, \citenamefont {Staykov}, \citenamefont {Kuan}, \citenamefont {Doneva},\ and\ \citenamefont {Yazadjiev}}]{Lam:2025ason}%
  \BibitemOpen
  \bibfield  {author} {\bibinfo {author} {\bibfnamefont {A.~T.-L.}\ \bibnamefont {Lam}}, \bibinfo {author} {\bibfnamefont {K.~V.}\ \bibnamefont {Staykov}}, \bibinfo {author} {\bibfnamefont {H.-J.}\ \bibnamefont {Kuan}}, \bibinfo {author} {\bibfnamefont {D.~D.}\ \bibnamefont {Doneva}},\ and\ \bibinfo {author} {\bibfnamefont {S.~S.}\ \bibnamefont {Yazadjiev}},\ }\bibfield  {title} {\bibinfo {title} {Axisymmetric stability of neutron stars as extreme rotators in massive scalar-tensor theory},\ }\href {https://doi.org/10.1103/PhysRevD.111.104030} {\bibfield  {journal} {\bibinfo  {journal} {Phys. Rev. D}\ }\textbf {\bibinfo {volume} {111}},\ \bibinfo {pages} {104030} (\bibinfo {year} {2025})}\BibitemShut {NoStop}%
\bibitem [{\citenamefont {Kuan}\ \emph {et~al.}(2021)\citenamefont {Kuan}, \citenamefont {Doneva},\ and\ \citenamefont {Yazadjiev}}]{Kuan:2021}%
  \BibitemOpen
  \bibfield  {author} {\bibinfo {author} {\bibfnamefont {H.-J.}\ \bibnamefont {Kuan}}, \bibinfo {author} {\bibfnamefont {D.~D.}\ \bibnamefont {Doneva}},\ and\ \bibinfo {author} {\bibfnamefont {S.~S.}\ \bibnamefont {Yazadjiev}},\ }\bibfield  {title} {\bibinfo {title} {Dynamical formation of scalarized black holes and neutron stars through stellar core collapse},\ }\href {https://doi.org/10.1103/PhysRevLett.127.161103} {\bibfield  {journal} {\bibinfo  {journal} {Phys. Rev. Lett.}\ }\textbf {\bibinfo {volume} {127}},\ \bibinfo {pages} {161103} (\bibinfo {year} {2021})}\BibitemShut {NoStop}%
\bibitem [{\citenamefont {Silva}\ \emph {et~al.}(2018{\natexlab{a}})\citenamefont {Silva}, \citenamefont {Sakstein}, \citenamefont {Gualtieri}, \citenamefont {Sotiriou},\ and\ \citenamefont {Berti}}]{Silva:2018ssb}%
  \BibitemOpen
  \bibfield  {author} {\bibinfo {author} {\bibfnamefont {H.~O.}\ \bibnamefont {Silva}}, \bibinfo {author} {\bibfnamefont {J.}~\bibnamefont {Sakstein}}, \bibinfo {author} {\bibfnamefont {L.}~\bibnamefont {Gualtieri}}, \bibinfo {author} {\bibfnamefont {T.~P.}\ \bibnamefont {Sotiriou}},\ and\ \bibinfo {author} {\bibfnamefont {E.}~\bibnamefont {Berti}},\ }\bibfield  {title} {\bibinfo {title} {Spontaneous scalarization of black holes and compact stars from a gauss-bonnet coupling},\ }\href {https://doi.org/10.1103/PhysRevLett.120.131104} {\bibfield  {journal} {\bibinfo  {journal} {Phys. Rev. Lett.}\ }\textbf {\bibinfo {volume} {120}},\ \bibinfo {pages} {131104} (\bibinfo {year} {2018}{\natexlab{a}})}\BibitemShut {NoStop}%
\bibitem [{\citenamefont {Herdeiro}\ \emph {et~al.}(2021{\natexlab{a}})\citenamefont {Herdeiro}, \citenamefont {Radu}, \citenamefont {Silva}, \citenamefont {Sotiriou},\ and\ \citenamefont {Yunes}}]{Herdeiro:2018}%
  \BibitemOpen
  \bibfield  {author} {\bibinfo {author} {\bibfnamefont {C.~A.~R.}\ \bibnamefont {Herdeiro}}, \bibinfo {author} {\bibfnamefont {E.}~\bibnamefont {Radu}}, \bibinfo {author} {\bibfnamefont {H.~O.}\ \bibnamefont {Silva}}, \bibinfo {author} {\bibfnamefont {T.~P.}\ \bibnamefont {Sotiriou}},\ and\ \bibinfo {author} {\bibfnamefont {N.}~\bibnamefont {Yunes}},\ }\bibfield  {title} {\bibinfo {title} {Spin-induced scalarized black holes},\ }\href {https://doi.org/10.1103/PhysRevLett.126.011103} {\bibfield  {journal} {\bibinfo  {journal} {Phys. Rev. Lett.}\ }\textbf {\bibinfo {volume} {126}},\ \bibinfo {pages} {011103} (\bibinfo {year} {2021}{\natexlab{a}})}\BibitemShut {NoStop}%
\bibitem [{\citenamefont {Li}\ \emph {et~al.}(2026)\citenamefont {Li}, \citenamefont {Sun},\ and\ \citenamefont {Yang}}]{Li:2026}%
  \BibitemOpen
  \bibfield  {author} {\bibinfo {author} {\bibfnamefont {L.}~\bibnamefont {Li}}, \bibinfo {author} {\bibfnamefont {Z.}~\bibnamefont {Sun}},\ and\ \bibinfo {author} {\bibfnamefont {F.-G.}\ \bibnamefont {Yang}},\ }\bibfield  {title} {\bibinfo {title} {Critical phenomenon inside asymptotically flat black holes with spontaneous scalarization},\ }\href {https://doi.org/10.1103/l4tf-kqyl} {\bibfield  {journal} {\bibinfo  {journal} {Phys. Rev. Lett.}\ }\textbf {\bibinfo {volume} {136}},\ \bibinfo {pages} {251402} (\bibinfo {year} {2026})}\BibitemShut {NoStop}%
\bibitem [{\citenamefont {Harada}(1998)}]{Harada:1998ge}%
  \BibitemOpen
  \bibfield  {author} {\bibinfo {author} {\bibfnamefont {T.}~\bibnamefont {Harada}},\ }\bibfield  {title} {\bibinfo {title} {{Neutron stars in scalar tensor theories of gravity and catastrophe theory}},\ }\href {https://doi.org/10.1103/PhysRevD.57.4802} {\bibfield  {journal} {\bibinfo  {journal} {Phys. Rev. D}\ }\textbf {\bibinfo {volume} {57}},\ \bibinfo {pages} {4802} (\bibinfo {year} {1998})},\ \Eprint {https://arxiv.org/abs/gr-qc/9801049} {arXiv:gr-qc/9801049} \BibitemShut {NoStop}%
\bibitem [{\citenamefont {Muniz}\ \emph {et~al.}(2025)\citenamefont {Muniz}, \citenamefont {Ortiz},\ and\ \citenamefont {Mendes}}]{Muniz:2025egq}%
  \BibitemOpen
  \bibfield  {author} {\bibinfo {author} {\bibfnamefont {J.~V.~M.}\ \bibnamefont {Muniz}}, \bibinfo {author} {\bibfnamefont {N.}~\bibnamefont {Ortiz}},\ and\ \bibinfo {author} {\bibfnamefont {R.~F.~P.}\ \bibnamefont {Mendes}},\ }\bibfield  {title} {\bibinfo {title} {Phase transition mechanism of spontaneous scalarization},\ }\href {https://doi.org/10.1103/qh47-qv76} {\bibfield  {journal} {\bibinfo  {journal} {Phys. Rev. D}\ }\textbf {\bibinfo {volume} {112}},\ \bibinfo {pages} {064037} (\bibinfo {year} {2025})},\ \Eprint {https://arxiv.org/abs/2503.11385} {arXiv:2503.11385 [gr-qc]} \BibitemShut {NoStop}%
\bibitem [{\citenamefont {{\"O}zinan}\ \emph {et~al.}(2026)\citenamefont {{\"O}zinan}, \citenamefont {{\"U}nl{\"u}t{\"u}rk},\ and\ \citenamefont {Ramazano{\u{g}}lu}}]{Ozinan:2026bne}%
  \BibitemOpen
  \bibfield  {author} {\bibinfo {author} {\bibfnamefont {M.}~\bibnamefont {{\"O}zinan}}, \bibinfo {author} {\bibfnamefont {K.~{\.I}.}\ \bibnamefont {{\"U}nl{\"u}t{\"u}rk}},\ and\ \bibinfo {author} {\bibfnamefont {F.~M.}\ \bibnamefont {Ramazano{\u{g}}lu}},\ }\bibfield  {title} {\bibinfo {title} {{Underlying mechanisms of phase transitions in scalar-tensor theories}},\ }\href {https://doi.org/10.1103/1brh-qzls} {\bibfield  {journal} {\bibinfo  {journal} {Phys. Rev. D}\ }\textbf {\bibinfo {volume} {114}},\ \bibinfo {pages} {024054} (\bibinfo {year} {2026})},\ \Eprint {https://arxiv.org/abs/2604.24867} {arXiv:2604.24867 [gr-qc]} \BibitemShut {NoStop}%
\bibitem [{\citenamefont {Huang}\ \emph {et~al.}(2025)\citenamefont {Huang}, \citenamefont {Kleihaus}, \citenamefont {Kunz}, \citenamefont {Lai}, \citenamefont {Radu},\ and\ \citenamefont {Zou}}]{Huang:2025dgc}%
  \BibitemOpen
  \bibfield  {author} {\bibinfo {author} {\bibfnamefont {H.}~\bibnamefont {Huang}}, \bibinfo {author} {\bibfnamefont {B.}~\bibnamefont {Kleihaus}}, \bibinfo {author} {\bibfnamefont {J.}~\bibnamefont {Kunz}}, \bibinfo {author} {\bibfnamefont {M.-Y.}\ \bibnamefont {Lai}}, \bibinfo {author} {\bibfnamefont {E.}~\bibnamefont {Radu}},\ and\ \bibinfo {author} {\bibfnamefont {D.-C.}\ \bibnamefont {Zou}},\ }\bibfield  {title} {\bibinfo {title} {{Phase transitions of boson stars in scalar-tensor theories}},\ }\href {https://doi.org/10.1103/8l73-dxxd} {\bibfield  {journal} {\bibinfo  {journal} {Phys. Rev. D}\ }\textbf {\bibinfo {volume} {112}},\ \bibinfo {pages} {124086} (\bibinfo {year} {2025})},\ \Eprint {https://arxiv.org/abs/2509.05202} {arXiv:2509.05202 [gr-qc]} \BibitemShut {NoStop}%
\bibitem [{\citenamefont {Guo}\ \emph {et~al.}(2026)\citenamefont {Guo}, \citenamefont {Liu},\ and\ \citenamefont {Myung}}]{Guo:2025flg}%
  \BibitemOpen
  \bibfield  {author} {\bibinfo {author} {\bibfnamefont {H.}~\bibnamefont {Guo}}, \bibinfo {author} {\bibfnamefont {H.}~\bibnamefont {Liu}},\ and\ \bibinfo {author} {\bibfnamefont {Y.~S.}\ \bibnamefont {Myung}},\ }\bibfield  {title} {\bibinfo {title} {{Scalar-hairy AdS black hole in the Einstein{\textendash}Maxwell-scalar theory: first-order phase transition with a critical point}},\ }\href {https://doi.org/10.1140/epjc/s10052-026-15407-7} {\bibfield  {journal} {\bibinfo  {journal} {Eur. Phys. J. C}\ }\textbf {\bibinfo {volume} {86}},\ \bibinfo {pages} {204} (\bibinfo {year} {2026})},\ \Eprint {https://arxiv.org/abs/2512.22433} {arXiv:2512.22433 [gr-qc]} \BibitemShut {NoStop}%
\bibitem [{\citenamefont {Herdeiro}\ \emph {et~al.}(2026)\citenamefont {Herdeiro}, \citenamefont {Huang}, \citenamefont {Kunz}, \citenamefont {Lai}, \citenamefont {Radu},\ and\ \citenamefont {Zou}}]{Herdeiro:2026}%
  \BibitemOpen
  \bibfield  {author} {\bibinfo {author} {\bibfnamefont {C.}~\bibnamefont {Herdeiro}}, \bibinfo {author} {\bibfnamefont {H.}~\bibnamefont {Huang}}, \bibinfo {author} {\bibfnamefont {J.}~\bibnamefont {Kunz}}, \bibinfo {author} {\bibfnamefont {M.-Y.}\ \bibnamefont {Lai}}, \bibinfo {author} {\bibfnamefont {E.}~\bibnamefont {Radu}},\ and\ \bibinfo {author} {\bibfnamefont {D.-C.}\ \bibnamefont {Zou}},\ }\bibfield  {title} {\bibinfo {title} {Phase structure of scalarized black holes in einstein-scalar-gauss-bonnet gravity},\ }\href {https://doi.org/10.1103/4k1l-jzvz} {\bibfield  {journal} {\bibinfo  {journal} {Phys. Rev. D}\ }\textbf {\bibinfo {volume} {113}},\ \bibinfo {pages} {124046} (\bibinfo {year} {2026})}\BibitemShut {NoStop}%
\bibitem [{\citenamefont {Nee}\ \emph {et~al.}(2025)\citenamefont {Nee}, \citenamefont {Lara}, \citenamefont {Pfeiffer},\ and\ \citenamefont {Vu}}]{PJNee:2025}%
  \BibitemOpen
  \bibfield  {author} {\bibinfo {author} {\bibfnamefont {P.~J.}\ \bibnamefont {Nee}}, \bibinfo {author} {\bibfnamefont {G.}~\bibnamefont {Lara}}, \bibinfo {author} {\bibfnamefont {H.~P.}\ \bibnamefont {Pfeiffer}},\ and\ \bibinfo {author} {\bibfnamefont {N.~L.}\ \bibnamefont {Vu}},\ }\bibfield  {title} {\bibinfo {title} {Quasistationary hair for binary black hole initial data in scalar gauss-bonnet gravity},\ }\href {https://doi.org/10.1103/PhysRevD.111.024061} {\bibfield  {journal} {\bibinfo  {journal} {Phys. Rev. D}\ }\textbf {\bibinfo {volume} {111}},\ \bibinfo {pages} {024061} (\bibinfo {year} {2025})}\BibitemShut {NoStop}%
\bibitem [{\citenamefont {Juli{\'e}}(2023)}]{Julie:2023ncq}%
  \BibitemOpen
  \bibfield  {author} {\bibinfo {author} {\bibfnamefont {F.-L.}\ \bibnamefont {Juli{\'e}}},\ }\href@noop {} {\bibinfo {title} {{Dynamical scalarization in Schwarzschild binary inspirals}}} (\bibinfo {year} {2023}),\ \Eprint {https://arxiv.org/abs/2312.16764} {arXiv:2312.16764 [gr-qc]} \BibitemShut {NoStop}%
\bibitem [{\citenamefont {Barausse}\ \emph {et~al.}(2013)\citenamefont {Barausse}, \citenamefont {Palenzuela}, \citenamefont {Ponce},\ and\ \citenamefont {Lehner}}]{Barausse:2012da}%
  \BibitemOpen
  \bibfield  {author} {\bibinfo {author} {\bibfnamefont {E.}~\bibnamefont {Barausse}}, \bibinfo {author} {\bibfnamefont {C.}~\bibnamefont {Palenzuela}}, \bibinfo {author} {\bibfnamefont {M.}~\bibnamefont {Ponce}},\ and\ \bibinfo {author} {\bibfnamefont {L.}~\bibnamefont {Lehner}},\ }\bibfield  {title} {\bibinfo {title} {{Neutron-star mergers in scalar-tensor theories of gravity}},\ }\href {https://doi.org/10.1103/PhysRevD.87.081506} {\bibfield  {journal} {\bibinfo  {journal} {Phys. Rev. D}\ }\textbf {\bibinfo {volume} {87}},\ \bibinfo {pages} {081506} (\bibinfo {year} {2013})},\ \Eprint {https://arxiv.org/abs/1212.5053} {arXiv:1212.5053 [gr-qc]} \BibitemShut {NoStop}%
\bibitem [{\citenamefont {Shibata}\ \emph {et~al.}(2014)\citenamefont {Shibata}, \citenamefont {Taniguchi}, \citenamefont {Okawa},\ and\ \citenamefont {Buonanno}}]{Shibata:2013pra}%
  \BibitemOpen
  \bibfield  {author} {\bibinfo {author} {\bibfnamefont {M.}~\bibnamefont {Shibata}}, \bibinfo {author} {\bibfnamefont {K.}~\bibnamefont {Taniguchi}}, \bibinfo {author} {\bibfnamefont {H.}~\bibnamefont {Okawa}},\ and\ \bibinfo {author} {\bibfnamefont {A.}~\bibnamefont {Buonanno}},\ }\bibfield  {title} {\bibinfo {title} {{Coalescence of binary neutron stars in a scalar-tensor theory of gravity}},\ }\href {https://doi.org/10.1103/PhysRevD.89.084005} {\bibfield  {journal} {\bibinfo  {journal} {Phys. Rev. D}\ }\textbf {\bibinfo {volume} {89}},\ \bibinfo {pages} {084005} (\bibinfo {year} {2014})},\ \Eprint {https://arxiv.org/abs/1310.0627} {arXiv:1310.0627 [gr-qc]} \BibitemShut {NoStop}%
\bibitem [{\citenamefont {Palenzuela}\ \emph {et~al.}(2014)\citenamefont {Palenzuela}, \citenamefont {Barausse}, \citenamefont {Ponce},\ and\ \citenamefont {Lehner}}]{Palenzuela:2013hsa}%
  \BibitemOpen
  \bibfield  {author} {\bibinfo {author} {\bibfnamefont {C.}~\bibnamefont {Palenzuela}}, \bibinfo {author} {\bibfnamefont {E.}~\bibnamefont {Barausse}}, \bibinfo {author} {\bibfnamefont {M.}~\bibnamefont {Ponce}},\ and\ \bibinfo {author} {\bibfnamefont {L.}~\bibnamefont {Lehner}},\ }\bibfield  {title} {\bibinfo {title} {{Dynamical scalarization of neutron stars in scalar-tensor gravity theories}},\ }\href {https://doi.org/10.1103/PhysRevD.89.044024} {\bibfield  {journal} {\bibinfo  {journal} {Phys. Rev. D}\ }\textbf {\bibinfo {volume} {89}},\ \bibinfo {pages} {044024} (\bibinfo {year} {2014})},\ \Eprint {https://arxiv.org/abs/1310.4481} {arXiv:1310.4481 [gr-qc]} \BibitemShut {NoStop}%
\bibitem [{\citenamefont {Dima}\ \emph {et~al.}(2020)\citenamefont {Dima}, \citenamefont {Barausse}, \citenamefont {Franchini},\ and\ \citenamefont {Sotiriou}}]{Dima:2020yac}%
  \BibitemOpen
  \bibfield  {author} {\bibinfo {author} {\bibfnamefont {A.}~\bibnamefont {Dima}}, \bibinfo {author} {\bibfnamefont {E.}~\bibnamefont {Barausse}}, \bibinfo {author} {\bibfnamefont {N.}~\bibnamefont {Franchini}},\ and\ \bibinfo {author} {\bibfnamefont {T.~P.}\ \bibnamefont {Sotiriou}},\ }\bibfield  {title} {\bibinfo {title} {{Spin-induced black hole spontaneous scalarization}},\ }\href {https://doi.org/10.1103/PhysRevLett.125.231101} {\bibfield  {journal} {\bibinfo  {journal} {Phys. Rev. Lett.}\ }\textbf {\bibinfo {volume} {125}},\ \bibinfo {pages} {231101} (\bibinfo {year} {2020})},\ \Eprint {https://arxiv.org/abs/2006.03095} {arXiv:2006.03095 [gr-qc]} \BibitemShut {NoStop}%
\bibitem [{\citenamefont {Herdeiro}\ \emph {et~al.}(2021{\natexlab{b}})\citenamefont {Herdeiro}, \citenamefont {Radu}, \citenamefont {Silva}, \citenamefont {Sotiriou},\ and\ \citenamefont {Yunes}}]{Herdeiro:2020wei}%
  \BibitemOpen
  \bibfield  {author} {\bibinfo {author} {\bibfnamefont {C.~A.~R.}\ \bibnamefont {Herdeiro}}, \bibinfo {author} {\bibfnamefont {E.}~\bibnamefont {Radu}}, \bibinfo {author} {\bibfnamefont {H.~O.}\ \bibnamefont {Silva}}, \bibinfo {author} {\bibfnamefont {T.~P.}\ \bibnamefont {Sotiriou}},\ and\ \bibinfo {author} {\bibfnamefont {N.}~\bibnamefont {Yunes}},\ }\bibfield  {title} {\bibinfo {title} {{Spin-induced scalarized black holes}},\ }\href {https://doi.org/10.1103/PhysRevLett.126.011103} {\bibfield  {journal} {\bibinfo  {journal} {Phys. Rev. Lett.}\ }\textbf {\bibinfo {volume} {126}},\ \bibinfo {pages} {011103} (\bibinfo {year} {2021}{\natexlab{b}})},\ \Eprint {https://arxiv.org/abs/2009.03904} {arXiv:2009.03904 [gr-qc]} \BibitemShut {NoStop}%
\bibitem [{\citenamefont {Elley}\ \emph {et~al.}(2022{\natexlab{a}})\citenamefont {Elley}, \citenamefont {Silva}, \citenamefont {Witek},\ and\ \citenamefont {Yunes}}]{Elley:2022ept}%
  \BibitemOpen
  \bibfield  {author} {\bibinfo {author} {\bibfnamefont {M.}~\bibnamefont {Elley}}, \bibinfo {author} {\bibfnamefont {H.~O.}\ \bibnamefont {Silva}}, \bibinfo {author} {\bibfnamefont {H.}~\bibnamefont {Witek}},\ and\ \bibinfo {author} {\bibfnamefont {N.}~\bibnamefont {Yunes}},\ }\bibfield  {title} {\bibinfo {title} {{Spin-induced dynamical scalarization, descalarization, and stealthness in scalar-Gauss-Bonnet gravity during a black hole coalescence}},\ }\href {https://doi.org/10.1103/PhysRevD.106.044018} {\bibfield  {journal} {\bibinfo  {journal} {Phys. Rev. D}\ }\textbf {\bibinfo {volume} {106}},\ \bibinfo {pages} {044018} (\bibinfo {year} {2022}{\natexlab{a}})},\ \Eprint {https://arxiv.org/abs/2205.06240} {arXiv:2205.06240 [gr-qc]} \BibitemShut {NoStop}%
\bibitem [{\citenamefont {Pardoe}\ and\ \citenamefont {Witek}(2026)}]{Pardoe:2026tcg}%
  \BibitemOpen
  \bibfield  {author} {\bibinfo {author} {\bibfnamefont {F.~C.~L.}\ \bibnamefont {Pardoe}}\ and\ \bibinfo {author} {\bibfnamefont {H.}~\bibnamefont {Witek}},\ }\href@noop {} {\bibinfo {title} {{Scalarization and descalarization in hyperbolic encounters of black holes}}} (\bibinfo {year} {2026}),\ \Eprint {https://arxiv.org/abs/2606.30865} {arXiv:2606.30865 [gr-qc]} \BibitemShut {NoStop}%
\bibitem [{\citenamefont {Capuano}\ \emph {et~al.}(2026)\citenamefont {Capuano}, \citenamefont {Sal\'o}, \citenamefont {Doneva}, \citenamefont {Yazadjiev},\ and\ \citenamefont {Barausse}}]{Capuano:2026}%
  \BibitemOpen
  \bibfield  {author} {\bibinfo {author} {\bibfnamefont {L.}~\bibnamefont {Capuano}}, \bibinfo {author} {\bibfnamefont {L.~A.}\ \bibnamefont {Sal\'o}}, \bibinfo {author} {\bibfnamefont {D.~D.}\ \bibnamefont {Doneva}}, \bibinfo {author} {\bibfnamefont {S.~S.}\ \bibnamefont {Yazadjiev}},\ and\ \bibinfo {author} {\bibfnamefont {E.}~\bibnamefont {Barausse}},\ }\bibfield  {title} {\bibinfo {title} {Dynamical hair growth in black hole binaries in einstein-scalar-gauss-bonnet gravity},\ }\href {https://doi.org/10.1103/hgfh-rwrw} {\bibfield  {journal} {\bibinfo  {journal} {Phys. Rev. D}\ }\textbf {\bibinfo {volume} {113}},\ \bibinfo {pages} {124038} (\bibinfo {year} {2026})}\BibitemShut {NoStop}%
\bibitem [{\citenamefont {Elley}\ \emph {et~al.}(2022{\natexlab{b}})\citenamefont {Elley}, \citenamefont {Silva}, \citenamefont {Witek},\ and\ \citenamefont {Yunes}}]{Elley:2022}%
  \BibitemOpen
  \bibfield  {author} {\bibinfo {author} {\bibfnamefont {M.}~\bibnamefont {Elley}}, \bibinfo {author} {\bibfnamefont {H.~O.}\ \bibnamefont {Silva}}, \bibinfo {author} {\bibfnamefont {H.}~\bibnamefont {Witek}},\ and\ \bibinfo {author} {\bibfnamefont {N.}~\bibnamefont {Yunes}},\ }\bibfield  {title} {\bibinfo {title} {Spin-induced dynamical scalarization, descalarization, and stealthness in scalar-gauss-bonnet gravity during a black hole coalescence},\ }\href {https://doi.org/10.1103/PhysRevD.106.044018} {\bibfield  {journal} {\bibinfo  {journal} {Phys. Rev. D}\ }\textbf {\bibinfo {volume} {106}},\ \bibinfo {pages} {044018} (\bibinfo {year} {2022}{\natexlab{b}})}\BibitemShut {NoStop}%
\bibitem [{Note1()}]{Note1}%
  \BibitemOpen
  \bibinfo {note} {One important exception is \protect \rev@citet {Kuan:2022oxs}, which considered a discontinuous transition at the maximum mass limit of stellar solutions, a distinct phenomenon from the later discover first-order scalarization picture we study here.}\BibitemShut {Stop}%
\bibitem [{\citenamefont {\"Ozel}\ and\ \citenamefont {Freire}(2016)}]{Ozel:2016oaf}%
  \BibitemOpen
  \bibfield  {author} {\bibinfo {author} {\bibfnamefont {F.}~\bibnamefont {\"Ozel}}\ and\ \bibinfo {author} {\bibfnamefont {P.}~\bibnamefont {Freire}},\ }\bibfield  {title} {\bibinfo {title} {{Masses, Radii, and the Equation of State of Neutron Stars}},\ }\href {https://doi.org/10.1146/annurev-astro-081915-023322} {\bibfield  {journal} {\bibinfo  {journal} {Ann. Rev. Astron. Astrophys.}\ }\textbf {\bibinfo {volume} {54}},\ \bibinfo {pages} {401} (\bibinfo {year} {2016})},\ \Eprint {https://arxiv.org/abs/1603.02698} {arXiv:1603.02698 [astro-ph.HE]} \BibitemShut {NoStop}%
\bibitem [{\citenamefont {Doneva}\ and\ \citenamefont {Yazadjiev}(2018)}]{Doneva:2017bvd}%
  \BibitemOpen
  \bibfield  {author} {\bibinfo {author} {\bibfnamefont {D.~D.}\ \bibnamefont {Doneva}}\ and\ \bibinfo {author} {\bibfnamefont {S.~S.}\ \bibnamefont {Yazadjiev}},\ }\bibfield  {title} {\bibinfo {title} {{New Gauss-Bonnet Black Holes with Curvature-Induced Scalarization in Extended Scalar-Tensor Theories}},\ }\href {https://doi.org/10.1103/PhysRevLett.120.131103} {\bibfield  {journal} {\bibinfo  {journal} {Phys. Rev. Lett.}\ }\textbf {\bibinfo {volume} {120}},\ \bibinfo {pages} {131103} (\bibinfo {year} {2018})},\ \Eprint {https://arxiv.org/abs/1711.01187} {arXiv:1711.01187 [gr-qc]} \BibitemShut {NoStop}%
\bibitem [{\citenamefont {Silva}\ \emph {et~al.}(2018{\natexlab{b}})\citenamefont {Silva}, \citenamefont {Sakstein}, \citenamefont {Gualtieri}, \citenamefont {Sotiriou},\ and\ \citenamefont {Berti}}]{Silva:2017uqg}%
  \BibitemOpen
  \bibfield  {author} {\bibinfo {author} {\bibfnamefont {H.~O.}\ \bibnamefont {Silva}}, \bibinfo {author} {\bibfnamefont {J.}~\bibnamefont {Sakstein}}, \bibinfo {author} {\bibfnamefont {L.}~\bibnamefont {Gualtieri}}, \bibinfo {author} {\bibfnamefont {T.~P.}\ \bibnamefont {Sotiriou}},\ and\ \bibinfo {author} {\bibfnamefont {E.}~\bibnamefont {Berti}},\ }\bibfield  {title} {\bibinfo {title} {{Spontaneous scalarization of black holes and compact stars from a Gauss-Bonnet coupling}},\ }\href {https://doi.org/10.1103/PhysRevLett.120.131104} {\bibfield  {journal} {\bibinfo  {journal} {Phys. Rev. Lett.}\ }\textbf {\bibinfo {volume} {120}},\ \bibinfo {pages} {131104} (\bibinfo {year} {2018}{\natexlab{b}})},\ \Eprint {https://arxiv.org/abs/1711.02080} {arXiv:1711.02080 [gr-qc]} \BibitemShut {NoStop}%
\bibitem [{\citenamefont {Andreou}\ \emph {et~al.}(2019)\citenamefont {Andreou}, \citenamefont {Franchini}, \citenamefont {Ventagli},\ and\ \citenamefont {Sotiriou}}]{Andreou:2019ikc}%
  \BibitemOpen
  \bibfield  {author} {\bibinfo {author} {\bibfnamefont {N.}~\bibnamefont {Andreou}}, \bibinfo {author} {\bibfnamefont {N.}~\bibnamefont {Franchini}}, \bibinfo {author} {\bibfnamefont {G.}~\bibnamefont {Ventagli}},\ and\ \bibinfo {author} {\bibfnamefont {T.~P.}\ \bibnamefont {Sotiriou}},\ }\bibfield  {title} {\bibinfo {title} {{Spontaneous scalarization in generalized scalar-tensor theory}},\ }\href {https://doi.org/10.1103/PhysRevD.99.124022} {\bibfield  {journal} {\bibinfo  {journal} {Phys. Rev.}\ }\textbf {\bibinfo {volume} {D99}},\ \bibinfo {pages} {124022} (\bibinfo {year} {2019})},\ \Eprint {https://arxiv.org/abs/1904.06365} {arXiv:1904.06365 [gr-qc]} \BibitemShut {NoStop}%
\bibitem [{Note2()}]{Note2}%
  \BibitemOpen
  \bibinfo {note} {Similar functions have been used in the context of nonlinear scalarization as well~\cite {Doneva:2021tvn}.}\BibitemShut {Stop}%
\bibitem [{Note3()}]{Note3}%
  \BibitemOpen
  \bibinfo {note} {We could easily construct a single parameter family of functions that has a vanishing quadratic term in the Taylor expansion for $A^4$, such as $A_\rho ^4(\varphi ) = 1 + 1/2 \tanh {(4\rho \varphi ^2)}$, which would prefer first-order scalarization similarly to Eq.~\protect \eqref {eq:conformal_factor} with $\gamma = 12\beta ^2$. This would explicitly demonstrate that the number of parameters in $A(\varphi )$ is not a direct measure of how complicated or fine tuned the theory is. However we chose to keep the current form in Eq.~\protect \eqref {eq:conformal_factor} that has the additional $\gamma $ parameter for ease of comparison to the existing literature.}\BibitemShut {Stop}%
\bibitem [{\citenamefont {Staykov}\ \emph {et~al.}(2026)\citenamefont {Staykov}, \citenamefont {Ramazano\u{g}lu}, \citenamefont {Doneva},\ and\ \citenamefont {Yazadjiev}}]{Staykov:2026}%
  \BibitemOpen
  \bibfield  {author} {\bibinfo {author} {\bibfnamefont {K.~V.}\ \bibnamefont {Staykov}}, \bibinfo {author} {\bibfnamefont {F.~M.}\ \bibnamefont {Ramazano\u{g}lu}}, \bibinfo {author} {\bibfnamefont {D.~D.}\ \bibnamefont {Doneva}},\ and\ \bibinfo {author} {\bibfnamefont {S.~S.}\ \bibnamefont {Yazadjiev}},\ }\href {https://arxiv.org/abs/2604.21298} {\bibinfo {title} {Phase transition structure of scalarized neutron stars: the effect of rotation and linear coupling}} (\bibinfo {year} {2026}),\ \Eprint {https://arxiv.org/abs/2604.21298} {arXiv:2604.21298 [gr-qc]} \BibitemShut {NoStop}%
\bibitem [{\citenamefont {Rosca-Mead}\ \emph {et~al.}(2020{\natexlab{a}})\citenamefont {Rosca-Mead}, \citenamefont {Moore}, \citenamefont {Sperhake}, \citenamefont {Agathos},\ and\ \citenamefont {Gerosa}}]{Rosca-Mead:2020bzt}%
  \BibitemOpen
  \bibfield  {author} {\bibinfo {author} {\bibfnamefont {R.}~\bibnamefont {Rosca-Mead}}, \bibinfo {author} {\bibfnamefont {C.~J.}\ \bibnamefont {Moore}}, \bibinfo {author} {\bibfnamefont {U.}~\bibnamefont {Sperhake}}, \bibinfo {author} {\bibfnamefont {M.}~\bibnamefont {Agathos}},\ and\ \bibinfo {author} {\bibfnamefont {D.}~\bibnamefont {Gerosa}},\ }\bibfield  {title} {\bibinfo {title} {{Structure of neutron stars in massive scalar-tensor gravity}},\ }\href {https://doi.org/10.3390/sym12091384} {\bibfield  {journal} {\bibinfo  {journal} {Symmetry}\ }\textbf {\bibinfo {volume} {12}},\ \bibinfo {pages} {1384} (\bibinfo {year} {2020}{\natexlab{a}})},\ \Eprint {https://arxiv.org/abs/2007.14429} {arXiv:2007.14429 [gr-qc]} \BibitemShut {NoStop}%
\bibitem [{\citenamefont {Read}\ \emph {et~al.}(2009)\citenamefont {Read}, \citenamefont {Markakis}, \citenamefont {Shibata}, \citenamefont {Uryu}, \citenamefont {Creighton},\ and\ \citenamefont {Friedman}}]{Read:2009yp}%
  \BibitemOpen
  \bibfield  {author} {\bibinfo {author} {\bibfnamefont {J.~S.}\ \bibnamefont {Read}}, \bibinfo {author} {\bibfnamefont {C.}~\bibnamefont {Markakis}}, \bibinfo {author} {\bibfnamefont {M.}~\bibnamefont {Shibata}}, \bibinfo {author} {\bibfnamefont {K.}~\bibnamefont {Uryu}}, \bibinfo {author} {\bibfnamefont {J.~D.~E.}\ \bibnamefont {Creighton}},\ and\ \bibinfo {author} {\bibfnamefont {J.~L.}\ \bibnamefont {Friedman}},\ }\bibfield  {title} {\bibinfo {title} {{Measuring the neutron star equation of state with gravitational wave observations}},\ }\href {https://doi.org/10.1103/PhysRevD.79.124033} {\bibfield  {journal} {\bibinfo  {journal} {Phys. Rev. D}\ }\textbf {\bibinfo {volume} {79}},\ \bibinfo {pages} {124033} (\bibinfo {year} {2009})},\ \Eprint {https://arxiv.org/abs/0901.3258} {arXiv:0901.3258 [gr-qc]} \BibitemShut {NoStop}%
\bibitem [{\citenamefont {Evans}\ \emph {et~al.}(2023)\citenamefont {Evans}, \citenamefont {Sturani}, \citenamefont {Vitale},\ and\ \citenamefont {Hall}}]{LIGO-T1500293}%
  \BibitemOpen
  \bibfield  {author} {\bibinfo {author} {\bibfnamefont {M.}~\bibnamefont {Evans}}, \bibinfo {author} {\bibfnamefont {R.}~\bibnamefont {Sturani}}, \bibinfo {author} {\bibfnamefont {S.}~\bibnamefont {Vitale}},\ and\ \bibinfo {author} {\bibfnamefont {E.}~\bibnamefont {Hall}},\ }\href@noop {} {\bibinfo {title} {Unofficial sensitivity curves for aligo, kagra, virgo, voyager, cosmic explorer, and einstein telescope}},\ \bibinfo {howpublished} {LIGO Document T1500293-v13} (\bibinfo {year} {2023})\BibitemShut {NoStop}%
\bibitem [{\citenamefont {Ramazano\u{g}lu}\ and\ \citenamefont {Pretorius}(2016)}]{Ramazanoglu:2016kul}%
  \BibitemOpen
  \bibfield  {author} {\bibinfo {author} {\bibfnamefont {F.~M.}\ \bibnamefont {Ramazano\u{g}lu}}\ and\ \bibinfo {author} {\bibfnamefont {F.}~\bibnamefont {Pretorius}},\ }\bibfield  {title} {\bibinfo {title} {{Spontaneous Scalarization with Massive Fields}},\ }\href {https://doi.org/10.1103/PhysRevD.93.064005} {\bibfield  {journal} {\bibinfo  {journal} {Phys. Rev.}\ }\textbf {\bibinfo {volume} {D93}},\ \bibinfo {pages} {064005} (\bibinfo {year} {2016})},\ \Eprint {https://arxiv.org/abs/1601.07475} {arXiv:1601.07475 [gr-qc]} \BibitemShut {NoStop}%
\bibitem [{\citenamefont {Tuna}\ \emph {et~al.}(2022)\citenamefont {Tuna}, \citenamefont {\"Unl\"ut\"urk},\ and\ \citenamefont {Ramazano\u{g}lu}}]{Tuna:2022qqr}%
  \BibitemOpen
  \bibfield  {author} {\bibinfo {author} {\bibfnamefont {S.}~\bibnamefont {Tuna}}, \bibinfo {author} {\bibfnamefont {K.~I.}\ \bibnamefont {\"Unl\"ut\"urk}},\ and\ \bibinfo {author} {\bibfnamefont {F.~M.}\ \bibnamefont {Ramazano\u{g}lu}},\ }\bibfield  {title} {\bibinfo {title} {{Constraining scalar-tensor theories using neutron star mass and radius measurements}},\ }\href {https://doi.org/10.1103/PhysRevD.105.124070} {\bibfield  {journal} {\bibinfo  {journal} {Phys. Rev. D}\ }\textbf {\bibinfo {volume} {105}},\ \bibinfo {pages} {124070} (\bibinfo {year} {2022})},\ \Eprint {https://arxiv.org/abs/2204.02138} {arXiv:2204.02138 [gr-qc]} \BibitemShut {NoStop}%
\bibitem [{\citenamefont {Kuan}\ \emph {et~al.}(2023)\citenamefont {Kuan}, \citenamefont {Van~Aelst}, \citenamefont {Lam},\ and\ \citenamefont {Shibata}}]{Kuan:2023hrh}%
  \BibitemOpen
  \bibfield  {author} {\bibinfo {author} {\bibfnamefont {H.-J.}\ \bibnamefont {Kuan}}, \bibinfo {author} {\bibfnamefont {K.}~\bibnamefont {Van~Aelst}}, \bibinfo {author} {\bibfnamefont {A.~T.-L.}\ \bibnamefont {Lam}},\ and\ \bibinfo {author} {\bibfnamefont {M.}~\bibnamefont {Shibata}},\ }\bibfield  {title} {\bibinfo {title} {{Binary neutron star mergers in massive scalar-tensor theory: Quasiequilibrium states and dynamical enhancement of the scalarization}},\ }\href {https://doi.org/10.1103/PhysRevD.108.064057} {\bibfield  {journal} {\bibinfo  {journal} {Phys. Rev. D}\ }\textbf {\bibinfo {volume} {108}},\ \bibinfo {pages} {064057} (\bibinfo {year} {2023})},\ \Eprint {https://arxiv.org/abs/2309.01709} {arXiv:2309.01709 [gr-qc]} \BibitemShut {NoStop}%
\bibitem [{\citenamefont {Gerosa}\ \emph {et~al.}(2016)\citenamefont {Gerosa}, \citenamefont {Sperhake},\ and\ \citenamefont {Ott}}]{Gerosa:2016fri}%
  \BibitemOpen
  \bibfield  {author} {\bibinfo {author} {\bibfnamefont {D.}~\bibnamefont {Gerosa}}, \bibinfo {author} {\bibfnamefont {U.}~\bibnamefont {Sperhake}},\ and\ \bibinfo {author} {\bibfnamefont {C.~D.}\ \bibnamefont {Ott}},\ }\bibfield  {title} {\bibinfo {title} {{Numerical simulations of stellar collapse in scalar-tensor theories of gravity}},\ }\href {https://doi.org/10.1088/0264-9381/33/13/135002} {\bibfield  {journal} {\bibinfo  {journal} {Class. Quant. Grav.}\ }\textbf {\bibinfo {volume} {33}},\ \bibinfo {pages} {135002} (\bibinfo {year} {2016})},\ \Eprint {https://arxiv.org/abs/1602.06952} {arXiv:1602.06952 [gr-qc]} \BibitemShut {NoStop}%
\bibitem [{\citenamefont {Berti}\ \emph {et~al.}(2021)\citenamefont {Berti}, \citenamefont {Collodel}, \citenamefont {Kleihaus},\ and\ \citenamefont {Kunz}}]{Berti:2020kgk}%
  \BibitemOpen
  \bibfield  {author} {\bibinfo {author} {\bibfnamefont {E.}~\bibnamefont {Berti}}, \bibinfo {author} {\bibfnamefont {L.~G.}\ \bibnamefont {Collodel}}, \bibinfo {author} {\bibfnamefont {B.}~\bibnamefont {Kleihaus}},\ and\ \bibinfo {author} {\bibfnamefont {J.}~\bibnamefont {Kunz}},\ }\bibfield  {title} {\bibinfo {title} {{Spin-induced black-hole scalarization in Einstein-scalar-Gauss-Bonnet theory}},\ }\href {https://doi.org/10.1103/PhysRevLett.126.011104} {\bibfield  {journal} {\bibinfo  {journal} {Phys. Rev. Lett.}\ }\textbf {\bibinfo {volume} {126}},\ \bibinfo {pages} {011104} (\bibinfo {year} {2021})},\ \Eprint {https://arxiv.org/abs/2009.03905} {arXiv:2009.03905 [gr-qc]} \BibitemShut {NoStop}%
\bibitem [{\citenamefont {Doneva}\ \emph {et~al.}(2013)\citenamefont {Doneva}, \citenamefont {Yazadjiev}, \citenamefont {Stergioulas},\ and\ \citenamefont {Kokkotas}}]{Doneva:2013qva}%
  \BibitemOpen
  \bibfield  {author} {\bibinfo {author} {\bibfnamefont {D.~D.}\ \bibnamefont {Doneva}}, \bibinfo {author} {\bibfnamefont {S.~S.}\ \bibnamefont {Yazadjiev}}, \bibinfo {author} {\bibfnamefont {N.}~\bibnamefont {Stergioulas}},\ and\ \bibinfo {author} {\bibfnamefont {K.~D.}\ \bibnamefont {Kokkotas}},\ }\bibfield  {title} {\bibinfo {title} {{Rapidly rotating neutron stars in scalar-tensor theories of gravity}},\ }\href {https://doi.org/10.1103/PhysRevD.88.084060} {\bibfield  {journal} {\bibinfo  {journal} {Phys. Rev. D}\ }\textbf {\bibinfo {volume} {88}},\ \bibinfo {pages} {084060} (\bibinfo {year} {2013})},\ \Eprint {https://arxiv.org/abs/1309.0605} {arXiv:1309.0605 [gr-qc]} \BibitemShut {NoStop}%
\bibitem [{\citenamefont {Sagert}\ \emph {et~al.}(2009)\citenamefont {Sagert}, \citenamefont {Fischer}, \citenamefont {Hempel}, \citenamefont {Pagliara}, \citenamefont {Schaffner-Bielich}, \citenamefont {Mezzacappa}, \citenamefont {Thielemann},\ and\ \citenamefont {Liebend\"orfer}}]{Sagert:2009}%
  \BibitemOpen
  \bibfield  {author} {\bibinfo {author} {\bibfnamefont {I.}~\bibnamefont {Sagert}}, \bibinfo {author} {\bibfnamefont {T.}~\bibnamefont {Fischer}}, \bibinfo {author} {\bibfnamefont {M.}~\bibnamefont {Hempel}}, \bibinfo {author} {\bibfnamefont {G.}~\bibnamefont {Pagliara}}, \bibinfo {author} {\bibfnamefont {J.}~\bibnamefont {Schaffner-Bielich}}, \bibinfo {author} {\bibfnamefont {A.}~\bibnamefont {Mezzacappa}}, \bibinfo {author} {\bibfnamefont {F.-K.}\ \bibnamefont {Thielemann}},\ and\ \bibinfo {author} {\bibfnamefont {M.}~\bibnamefont {Liebend\"orfer}},\ }\bibfield  {title} {\bibinfo {title} {Signals of the qcd phase transition in core-collapse supernovae},\ }\href {https://doi.org/10.1103/PhysRevLett.102.081101} {\bibfield  {journal} {\bibinfo  {journal} {Phys. Rev. Lett.}\ }\textbf {\bibinfo {volume} {102}},\ \bibinfo {pages} {081101} (\bibinfo {year} {2009})}\BibitemShut {NoStop}%
\bibitem [{\citenamefont {Shao}(2019)}]{Shao:2019gjj}%
  \BibitemOpen
  \bibfield  {author} {\bibinfo {author} {\bibfnamefont {L.}~\bibnamefont {Shao}},\ }\bibfield  {title} {\bibinfo {title} {{Degeneracy in Studying the Supranuclear Equation of State and Modified Gravity with Neutron Stars}},\ }\href {https://doi.org/10.1063/1.5117806} {\bibfield  {journal} {\bibinfo  {journal} {AIP Conf. Proc.}\ }\textbf {\bibinfo {volume} {2127}},\ \bibinfo {pages} {020016} (\bibinfo {year} {2019})},\ \Eprint {https://arxiv.org/abs/1901.07546} {arXiv:1901.07546 [gr-qc]} \BibitemShut {NoStop}%
\bibitem [{\citenamefont {Most}\ \emph {et~al.}(2019)\citenamefont {Most}, \citenamefont {Papenfort}, \citenamefont {Dexheimer}, \citenamefont {Hanauske}, \citenamefont {Schramm}, \citenamefont {St\"ocker},\ and\ \citenamefont {Rezzolla}}]{Most:2019}%
  \BibitemOpen
  \bibfield  {author} {\bibinfo {author} {\bibfnamefont {E.~R.}\ \bibnamefont {Most}}, \bibinfo {author} {\bibfnamefont {L.~J.}\ \bibnamefont {Papenfort}}, \bibinfo {author} {\bibfnamefont {V.}~\bibnamefont {Dexheimer}}, \bibinfo {author} {\bibfnamefont {M.}~\bibnamefont {Hanauske}}, \bibinfo {author} {\bibfnamefont {S.}~\bibnamefont {Schramm}}, \bibinfo {author} {\bibfnamefont {H.}~\bibnamefont {St\"ocker}},\ and\ \bibinfo {author} {\bibfnamefont {L.}~\bibnamefont {Rezzolla}},\ }\bibfield  {title} {\bibinfo {title} {Signatures of quark-hadron phase transitions in general-relativistic neutron-star mergers},\ }\href {https://doi.org/10.1103/PhysRevLett.122.061101} {\bibfield  {journal} {\bibinfo  {journal} {Phys. Rev. Lett.}\ }\textbf {\bibinfo {volume} {122}},\ \bibinfo {pages} {061101} (\bibinfo {year} {2019})}\BibitemShut {NoStop}%
\bibitem [{\citenamefont {Cao}\ and\ \citenamefont {Chen}(2026)}]{Cao:2026dwk}%
  \BibitemOpen
  \bibfield  {author} {\bibinfo {author} {\bibfnamefont {Z.}~\bibnamefont {Cao}}\ and\ \bibinfo {author} {\bibfnamefont {L.-W.}\ \bibnamefont {Chen}},\ }\href@noop {} {\bibinfo {title} {{On the Possibility of a Strong First-Order Phase Transition in Neutron Stars}}} (\bibinfo {year} {2026}),\ \Eprint {https://arxiv.org/abs/2606.06378} {arXiv:2606.06378 [nucl-th]} \BibitemShut {NoStop}%
\bibitem [{\citenamefont {Salgado}\ \emph {et~al.}(1998)\citenamefont {Salgado}, \citenamefont {Sudarsky},\ and\ \citenamefont {Nucamendi}}]{Salgado:1998sg}%
  \BibitemOpen
  \bibfield  {author} {\bibinfo {author} {\bibfnamefont {M.}~\bibnamefont {Salgado}}, \bibinfo {author} {\bibfnamefont {D.}~\bibnamefont {Sudarsky}},\ and\ \bibinfo {author} {\bibfnamefont {U.}~\bibnamefont {Nucamendi}},\ }\bibfield  {title} {\bibinfo {title} {{On spontaneous scalarization}},\ }\href {https://doi.org/10.1103/PhysRevD.58.124003} {\bibfield  {journal} {\bibinfo  {journal} {Phys. Rev. D}\ }\textbf {\bibinfo {volume} {58}},\ \bibinfo {pages} {124003} (\bibinfo {year} {1998})},\ \Eprint {https://arxiv.org/abs/gr-qc/9806070} {arXiv:gr-qc/9806070} \BibitemShut {NoStop}%
\bibitem [{\citenamefont {Taniguchi}\ \emph {et~al.}(2015)\citenamefont {Taniguchi}, \citenamefont {Shibata},\ and\ \citenamefont {Buonanno}}]{Taniguchi:2014fqa}%
  \BibitemOpen
  \bibfield  {author} {\bibinfo {author} {\bibfnamefont {K.}~\bibnamefont {Taniguchi}}, \bibinfo {author} {\bibfnamefont {M.}~\bibnamefont {Shibata}},\ and\ \bibinfo {author} {\bibfnamefont {A.}~\bibnamefont {Buonanno}},\ }\bibfield  {title} {\bibinfo {title} {{Quasiequilibrium sequences of binary neutron stars undergoing dynamical scalarization}},\ }\href {https://doi.org/10.1103/PhysRevD.91.024033} {\bibfield  {journal} {\bibinfo  {journal} {Phys. Rev. D}\ }\textbf {\bibinfo {volume} {91}},\ \bibinfo {pages} {024033} (\bibinfo {year} {2015})},\ \Eprint {https://arxiv.org/abs/1410.0738} {arXiv:1410.0738 [gr-qc]} \BibitemShut {NoStop}%
\bibitem [{\citenamefont {Khalil}\ \emph {et~al.}(2019)\citenamefont {Khalil}, \citenamefont {Sennett}, \citenamefont {Steinhoff},\ and\ \citenamefont {Buonanno}}]{Khalil:2019wyy}%
  \BibitemOpen
  \bibfield  {author} {\bibinfo {author} {\bibfnamefont {M.}~\bibnamefont {Khalil}}, \bibinfo {author} {\bibfnamefont {N.}~\bibnamefont {Sennett}}, \bibinfo {author} {\bibfnamefont {J.}~\bibnamefont {Steinhoff}},\ and\ \bibinfo {author} {\bibfnamefont {A.}~\bibnamefont {Buonanno}},\ }\bibfield  {title} {\bibinfo {title} {{Theory-agnostic framework for dynamical scalarization of compact binaries}},\ }\href {https://doi.org/10.1103/PhysRevD.100.124013} {\bibfield  {journal} {\bibinfo  {journal} {Phys. Rev. D}\ }\textbf {\bibinfo {volume} {100}},\ \bibinfo {pages} {124013} (\bibinfo {year} {2019})},\ \Eprint {https://arxiv.org/abs/1906.08161} {arXiv:1906.08161 [gr-qc]} \BibitemShut {NoStop}%
\bibitem [{\citenamefont {Doneva}\ and\ \citenamefont {Yazadjiev}(2022)}]{Doneva:2021tvn}%
  \BibitemOpen
  \bibfield  {author} {\bibinfo {author} {\bibfnamefont {D.~D.}\ \bibnamefont {Doneva}}\ and\ \bibinfo {author} {\bibfnamefont {S.~S.}\ \bibnamefont {Yazadjiev}},\ }\bibfield  {title} {\bibinfo {title} {{Beyond the spontaneous scalarization: New fully nonlinear mechanism for the formation of scalarized black holes and its dynamical development}},\ }\href {https://doi.org/10.1103/PhysRevD.105.L041502} {\bibfield  {journal} {\bibinfo  {journal} {Phys. Rev. D}\ }\textbf {\bibinfo {volume} {105}},\ \bibinfo {pages} {L041502} (\bibinfo {year} {2022})},\ \Eprint {https://arxiv.org/abs/2107.01738} {arXiv:2107.01738 [gr-qc]} \BibitemShut {NoStop}%
\bibitem [{\citenamefont {O'Connor}\ and\ \citenamefont {Ott}(2010)}]{OConnor:2009vw}%
  \BibitemOpen
  \bibfield  {author} {\bibinfo {author} {\bibfnamefont {E.}~\bibnamefont {O'Connor}}\ and\ \bibinfo {author} {\bibfnamefont {C.~D.}\ \bibnamefont {Ott}},\ }\bibfield  {title} {\bibinfo {title} {{A New Open-Source Code for Spherically-Symmetric Stellar Collapse to Neutron Stars and Black Holes}},\ }\href {https://doi.org/10.1088/0264-9381/27/11/114103} {\bibfield  {journal} {\bibinfo  {journal} {Class.Quant.Grav.}\ }\textbf {\bibinfo {volume} {27}},\ \bibinfo {pages} {114103} (\bibinfo {year} {2010})},\ \Eprint {https://arxiv.org/abs/0912.2393} {arXiv:0912.2393 [astro-ph.HE]} \BibitemShut {NoStop}%
\bibitem [{\citenamefont {Rosca-Mead}\ \emph {et~al.}(2020{\natexlab{b}})\citenamefont {Rosca-Mead}, \citenamefont {Sperhake}, \citenamefont {Moore}, \citenamefont {Agathos}, \citenamefont {Gerosa},\ and\ \citenamefont {Ott}}]{Rosca-Mosca:2020collapse}%
  \BibitemOpen
  \bibfield  {author} {\bibinfo {author} {\bibfnamefont {R.}~\bibnamefont {Rosca-Mead}}, \bibinfo {author} {\bibfnamefont {U.}~\bibnamefont {Sperhake}}, \bibinfo {author} {\bibfnamefont {C.~J.}\ \bibnamefont {Moore}}, \bibinfo {author} {\bibfnamefont {M.}~\bibnamefont {Agathos}}, \bibinfo {author} {\bibfnamefont {D.}~\bibnamefont {Gerosa}},\ and\ \bibinfo {author} {\bibfnamefont {C.~D.}\ \bibnamefont {Ott}},\ }\bibfield  {title} {\bibinfo {title} {Core collapse in massive scalar-tensor gravity},\ }\href {https://doi.org/10.1103/PhysRevD.102.044010} {\bibfield  {journal} {\bibinfo  {journal} {Phys. Rev. D}\ }\textbf {\bibinfo {volume} {102}},\ \bibinfo {pages} {044010} (\bibinfo {year} {2020}{\natexlab{b}})}\BibitemShut {NoStop}%
\bibitem [{\citenamefont {Mendes}\ and\ \citenamefont {Ortiz}(2016)}]{Mendes:2016fby}%
  \BibitemOpen
  \bibfield  {author} {\bibinfo {author} {\bibfnamefont {R.~F.~P.}\ \bibnamefont {Mendes}}\ and\ \bibinfo {author} {\bibfnamefont {N.}~\bibnamefont {Ortiz}},\ }\bibfield  {title} {\bibinfo {title} {{Highly compact neutron stars in scalar-tensor theories of gravity: Spontaneous scalarization versus gravitational collapse}},\ }\href {https://doi.org/10.1103/PhysRevD.93.124035} {\bibfield  {journal} {\bibinfo  {journal} {Phys. Rev.}\ }\textbf {\bibinfo {volume} {D93}},\ \bibinfo {pages} {124035} (\bibinfo {year} {2016})},\ \Eprint {https://arxiv.org/abs/1604.04175} {arXiv:1604.04175 [gr-qc]} \BibitemShut {NoStop}%
\bibitem [{\citenamefont {Teukolsky}(2000)}]{Teukolsky:2000}%
  \BibitemOpen
  \bibfield  {author} {\bibinfo {author} {\bibfnamefont {S.~A.}\ \bibnamefont {Teukolsky}},\ }\bibfield  {title} {\bibinfo {title} {Stability of the iterated crank-nicholson method in numerical relativity},\ }\href {https://doi.org/10.1103/PhysRevD.61.087501} {\bibfield  {journal} {\bibinfo  {journal} {Phys. Rev. D}\ }\textbf {\bibinfo {volume} {61}},\ \bibinfo {pages} {087501} (\bibinfo {year} {2000})}\BibitemShut {NoStop}%
\end{thebibliography}%

\hfill\\

\onecolumngrid
\begin{center}
	\textbf{\large End Matter}
\end{center}
\twocolumngrid
\setcounter{equation}{0}
\setcounter{figure}{0}
\setcounter{table}{0}

\renewcommand{\thefigure}{E\arabic{figure}}
\setcounter{figure}{0}

\renewcommand\theequation{E\arabic{equation}}

\noindent {\bf \em Numerical Detail:} The stellar evolution is performed with a scalar-tensor extension of the GR1D code~\cite{OConnor:2009vw}. GR1D was originally developed for spherically symmetric relativistic hydrodynamics in general relativity and has subsequently been generalized to scalar-tensor gravity. We adopt the
same basic numerical formulation as in Secs.~II--III of \textcite{Rosca-Mosca:2020collapse}. The matter equations are written in flux-conservative form using the conserved variables, and the scalar field is evolved as a first-order system using the variables $\eta$ and $\psi$ defined in Eq.~(17) of \textcite{Rosca-Mosca:2020collapse}, and the scalar-tensor evolution system is given in Eqs.~(18)--(29) therein. The hydrodynamic sector is evolved with a finite-volume high-resolution shock-capturing scheme, while the metric and scalar equations are updated consistently with the spherical polar-slicing, radial-gauge formulation. Berger-Oliger type numerical dissipation is implemented for scalar variables following \textcite{Gerosa:2016fri}, with the same coefficient of $\mathcal{D} = 2$. However, our EOS implementation differs from the original core collapse code as we do not consider thermal effects.

\begin{figure}
    \centering
    \includegraphics[width=\columnwidth]{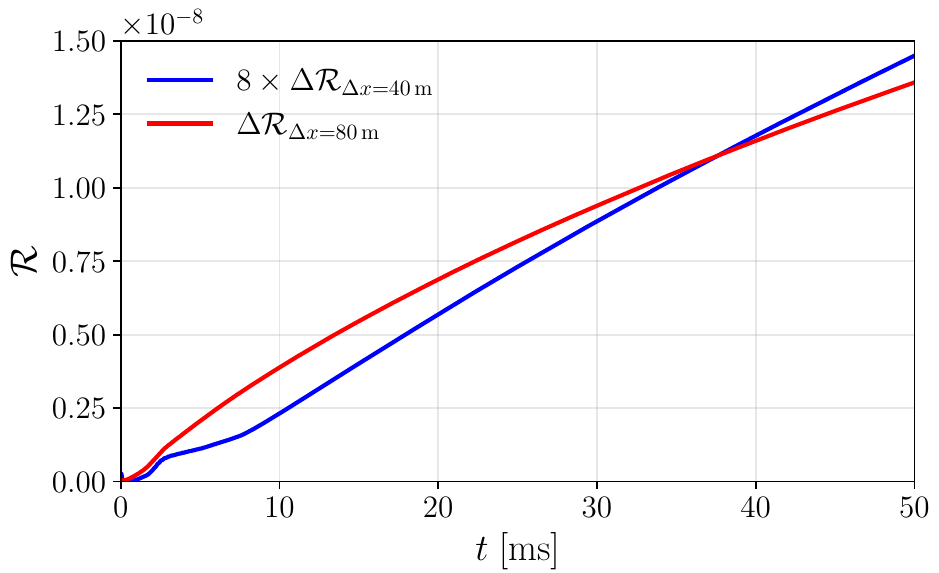}
    \caption{Convergence of the scalar residual $\mathcal{C}_{\varphi}$ in the $L_1$ norm.}
    \label{fig:scalar_residual}
\end{figure}

The scalar sector of our evolution system introduces a constraint associated with the first-order reduction of the scalar-field equation. Since we independently evolve the scalar field and its first space derivative, their difference allows us to test convergence of our scalar modification. Following \textcite{Mendes:2016fby},  we show the increase in the $L_1$ norm of the scalar residual during the time evolution, $ \Delta\mathcal{R}_{\varphi}(t)=\|\mathcal{C}_{\varphi}\|_1(t) - \|\mathcal{C}_{\varphi}\|_1(t= 0)$ in Figure~\ref{fig:scalar_residual}. We subtract the initial residual because the initial data are interpolated from a much higher-resolution grid, which introduces a small interpolation error at the beginning of the simulation. The lower resolution uses a uniform grid spacing of $\Delta r = 40\,\mathrm{m}$ up to $r = 32\,\mathrm{km}$, followed by a logarithmically spaced outer grid extending to $r = 5\times10^{4}\,\mathrm{km}$, with $4000$ grid points in total. The higher resolution halves the grid spacing, and the comparison shows approximately third order convergence as is expected from our third order Runge-Kutta time stepping.

\noindent {\bf \em Wave Propagation and Extraction:}
The scalar waveform is extracted from the GR1D simulation at a radius \(r_{\rm ext}\) where the space-time is approximately flat. Outside this radius, the scalar dynamics reduces to the flat-space massive Klein--Gordon equation. Defining
\begin{equation}
    u(t,r) \equiv r\varphi(t,r),
\end{equation}
the scalar field equation~\eqref{eq:scalar_equation} approximately becomes
\begin{equation} \label{eq: massive_klein}
    \partial_t^2 u
    -
    \partial_r^2 u
    +
    (2\pi f_*)^2 u
    =
    0 
\end{equation}
We need to evolve the scalar wave to a far-zone radius compared to the Compton wavelength of the scalar field, allowing the evanescent modes to decay, before taking the Fourier transform and propagating the signal to kiloparsec-scale
distances. Introducing the retarded time $t_r \equiv t-r$, we rewrite Eq.~\eqref{eq: massive_klein} as
\begin{equation}
    2\,\partial_{t_r}\partial_r u
    -\partial_r^2 u
    +\omega_*^2 u = 0 .
\end{equation}
We then introduce $\Pi \equiv \partial_{t_r}u$, and solve the resulting first order system with a method-of-lines discretization~\cite{Rosca-Mead:2020bzt}. The retarded-time update of $u$ is performed with an \textit{iterated Crank--Nicolson scheme}~\cite{Teukolsky:2000}. This method provides a suitable setup for resolving waves going near the speed of light~\cite{Rosca-Mead:2020bzt}. The inner boundary data for this flat-space propagation are supplied by the GR1D extraction, $ u(t_r,r_{\rm ext}) = r_{\rm ext}\varphi(t,r_{\rm ext}) $. Finally, we take the fourier transform at $2\times10^5 \, \mathrm{km}$ and follow \textcite{Sperhake:2017itk} for the remaining kpc scale propagation.

\noindent {\bf \em Accretion Model:}
The gaussian matter cloud absorbed by the star has a width of $\sigma=1.5\, \mathrm{km}$ and a peak density of $5\times 10^{-6}$ times the stellar central rest energy density. It increases the baryon mass of the star by $ \simeq 1.47 \times 10^{-5}\,M_\odot$. Although this accretion model is not fully realistic, its influence on the resulting GW signal and the first-order transition is negligible. We verified that the accretion merely triggers the transition and does not directly affect the astrophysical signals extracted in this study by varying the peak amplitude of the Gaussian. Increasing the accretion amplitude by a factor of $1000$ to match the same accretion rate as \textcite{Kuan:2023hrh} yields essentially the same GW signal as that shown in Figure~\ref{fig:sensitivity}.

\end{document}